%% file: beam_to_solid.tex
\RequirePackage{fix-cm}
\documentclass[twocolumn, envcountsect]{svjour3}          
\smartqed  
\usepackage{graphicx}
\usepackage{mathptmx}      
\input{tex/packages.tex}
\input{tex/shortcuts.tex}

\journalname{}

\begin{document}

\title{A mortar-type finite element approach for embedding 1D beams into 3D solid volumes}


\author{
	Ivo Steinbrecher\and
	Matthias Mayr\and
	Maximilian J. Grill\and
	Johannes Kremheller\and
	Christoph Meier\and
	Alexander Popp
}


\institute{
	I. Steinbrecher, M. Mayr, A.Popp\at
	Institute for Mathematics and Computer-Based Simulation,\\
    University of the Bundeswehr Munich,\\
    Werner-Heisenberg-Weg 39, D-85577 Neubiberg, Germany\\
    \email{ivo.steinbrecher@unibw.de}
    \and
    M. J. Grill, J. Kremheller, C. Meier\at
   	Institute for Computational Mechanics,\\
   	Technical University of Munich,\\
   	Boltzmannstrasse 15, D-85748 Garching b. München, Germany
}

\date{Received: date / Accepted: date}

\maketitle

\input{tex/abstract.tex}
\input{tex/introduction.tex}
\input{tex/problem_definition.tex}
\input{tex/discretization.tex}

\input{tex/examples.tex}
\input{tex/conclusions.tex}

%
%

\bibliographystyle{spmpsci}      
\bibliography{beam_to_solid}   

\end{document}

%% file: tex/packages.tex
\usepackage[T1]{fontenc}
\usepackage[utf8]{inputenc}
\usepackage{newtxtext,newtxmath}

\usepackage{xspace}

\usepackage{units}

\usepackage{todonotes}
\usepackage{color}

\usepackage{subfigure}

\usepackage{amsmath}
\usepackage{amssymb}

\usepackage{booktabs}

\usepackage{cite}

\usepackage{siunitx}

%% file: tex/shortcuts.tex
\newcommand{\R}[1]{\mathbb{R}^{#1}}
\renewcommand{\L}[1]{L_{#1}}
\newcommand{\C}[1]{\ensuremath{C^{#1}}}
\newcommand{\SO}{SO^3}

\newcommand{\intd}[1]{\mathrm{d}{#1}}

\newcommand{\tns}[1]{\underline{\boldsymbol{#1}}}
\newcommand{\tnss}[1]{\boldsymbol{#1}}
\newcommand{\tnssI}{\tnss{I}}
\newcommand{\tnsO}{\tns{0}}
\newcommand{\tr}{^{\mathrm T}}

\newcommand{\vv}[1]{\boldsymbol{#1}}
\newcommand{\vvO}{\vv{0}}
\newcommand{\mat}[1]{\boldsymbol{#1}}
\newcommand{\matO}{\mat{0}}
\newcommand{\matI}{\mat{I}}
\newcommand{\norm}[1]{\left\|#1\right\|}

\newcommand{\placeholder}{(\cdot)}

\newcommand{\br}[1]{\left(#1\right)}

\newcommand{\dirac}{\delta}

\newcommand{\lininline}[2]{\partial #1 / \partial #2}

\newcommand{\order}[1]{\mathcal O(#1)}

\newcommand{\ie}{i.e.~}
\newcommand{\eg}{e.g.~}
\newcommand{\cf}{cf.~}

\newcommand{\sts}{solid-to-solid\xspace}

\newcommand{\bis}{beam-in-solid\xspace}

\newcommand{\btsvc}{beam-to-solid volume coupling\xspace}
\newcommand{\Btsvc}{Beam-to-solid volume coupling\xspace}
\newcommand{\btssc}{beam-to-solid surface coupling\xspace}

\newcommand{\gptslong}{Gauss-point-to-segment\xspace}
\newcommand{\gpts}{GPTS\xspace}

\newcommand{\sr}{Simo--Reissner\xspace}
\newcommand{\kl}{Kirchhoff--Love\xspace}

\newcommand{\Tf}{Torsion-free\xspace}
\newcommand{\tf}{torsion-free\xspace}
\newcommand{\cs}[1][]{cross-section#1\xspace}

\renewcommand{\hex}[1]{\emph{hex#1}\xspace}
\newcommand{\xfem}{XFEM\xspace}

\newcommand{\ex}{\tns{e}_1}
\newcommand{\ey}{\tns{e}_2}
\newcommand{\ez}{\tns{e}_3}

\newcommand{\osolid}{\Omega^S}
\newcommand{\osolidO}{\osolid_0}
\newcommand{\osolidt}{\osolid}
\newcommand{\osolidOp}{\partial\osolidO}

\newcommand{\obeam}{\Omega^B}
\newcommand{\obeamLO}{\Omega^B_{L,0}}
\newcommand{\obeamL}{\Omega^B_L}
\newcommand{\obeamO}{\obeam_0}

\newcommand{\obeamOp}{\partial\obeamO}
\newcommand{\beamlength}{L}

\newcommand{\gcouplingf}{\Gamma_{c}^\twotothree}
\newcommand{\gcouplingl}{\Gamma_{c}^\onetothree}

\newcommand{\Gsolid}{\Gamma}
\newcommand{\Gsolidu}{\Gsolid_u}
\newcommand{\Gsolids}{\Gsolid_\sigma}

\newcommand{\gsolid}{\gamma}
\newcommand{\gsolidu}{\gsolid_u}
\newcommand{\gsolids}{\gsolid_\sigma}

\newcommand{\Xsolid}{\tns{X}^S}
\newcommand{\xsolid}{\tns{x}^S}
\newcommand{\usolid}{\tns{u}^S}
\newcommand{\dusolid}{\delta \usolid}

\newcommand{\sbeam}{s}
\newcommand{\Xbeam}{\tns{X}^B}
\newcommand{\xbeam}{\tns{x}^B}
\newcommand{\ubeam}{\tns{u}^B}
\newcommand{\ubeamr}{\tns{u}^B_r}
\newcommand{\dubeam}{\delta \ubeam}
\newcommand{\dubeamr}{\delta \ubeamr}
\newcommand{\rbeam}{\tns{r}}
\newcommand{\rbeamO}{\tns{r}_0}
\newcommand{\triad}{\tnss{\Lambda}}
\newcommand{\g}[1]{\tns{g}_{#1}}
\newcommand{\csa}{\alpha}
\newcommand{\csb}{\beta}
\newcommand{\rotvec}{\tns{\psi}}
\newcommand{\twist}{\varphi}
\newcommand{\frenet}{\tns{\kappa}}

\newcommand{\Spk}{\tnss{S}}
\newcommand{\F}{\tnss{F}}
\newcommand{\E}{\tnss{E}}
\newcommand{\dE}{\delta \E}

\newcommand{\VO}{V_0}
\newcommand{\dV}{\intd{\VO}}
\newcommand{\AO}{A_0}
\newcommand{\dA}{\intd{\AO}}

\newcommand{\dWs}{\delta W^S}
\newcommand{\dWb}{\delta W^B}

\newcommand{\dWints}{\delta W^S_{\mathrm{int}}}

\newcommand{\dWexts}{\delta W^S_{\mathrm{ext}}}

\newcommand{\onetothree}{\text{1D-3D}}
\newcommand{\twotothree}{\text{2D-3D}}
\newcommand{\dWmt}{\delta W_{c}}
\newcommand{\dWlambda}{\delta W_{\lambda}}
\newcommand{\dWmtl}{\dWmt^{\onetothree}}
\newcommand{\dWmtf}{\dWmt^{\twotothree}}
\newcommand{\dWmth}{\delta {{W}_{c,h}}}
\newcommand{\dWlambdal}{\dWlambda^{\onetothree}}
\newcommand{\dWlambdaf}{\dWlambda^{\twotothree}}
\newcommand{\dWlambdah}{\delta {{W}_{\lambda,h}}}

\newcommand{\lagrange}{\tns{\lambda}}
\newcommand{\lagrangel}{\tns{\lambda}^{\onetothree}}
\newcommand{\lagrangef}{\tns{\lambda}^{\twotothree}}

\newcommand{\dlagrangel}{\delta \lagrange^{\onetothree}}
\newcommand{\dlagrangef}{\delta \lagrange^{\twotothree}}

\newcommand{\esolidName}{e}
\newcommand{\esolid}{{(\esolidName)}}

\newcommand{\nsolid}{n_S}
\newcommand{\nbeam}{n_B}
\newcommand{\nlagrange}{n_\lambda}

\newcommand{\indexsolid}{k}
\newcommand{\indexbeam}{l}
\newcommand{\indexlagrange}{j}
\newcommand{\sumsolid}[1]{\sum_{\indexsolid=1}^{\nsolid}{#1}}
\newcommand{\sumbeam}[1]{\sum_{\indexbeam=1}^{\nbeam}{#1}}
\newcommand{\sumlagrange}[1]{\sum_{\indexlagrange=1}^{\nlagrange}{#1}}

\newcommand{\Xsolidh}{{\tns{X}^S_h}}
\newcommand{\usolidh}{{\tns{u}^S_h}}
\newcommand{\usolidhref}{{\tns{u}^S_{ref}}}

\newcommand{\ubeamrh}{{\tns{u}^B_{r,h}}}

\newcommand{\ubeamrhref}{{\tns{u}^B_{r,ref}}}

\newcommand{\rbeamOh}{\tns{r}_{0,h}}

\newcommand{\jacobian}{\left\| \partial \rbeamOh / \partial s \right\|}

\newcommand{\lagrangeh}{\tns{\lambda}_{h}}

\newcommand{\qsolid}{{\vv{d}^S}}

\newcommand{\qsolide}{\vv{d}^{S}_{\indexsolid}}
\newcommand{\qxsolide}{\vv{x}^{S}_{\indexsolid}}
\newcommand{\qbeam}{\vv{d}^B}

\newcommand{\qbeame}{\vv{d}^{B}_{\indexbeam}}
\newcommand{\qbeamer}{\vv{d}^{B,r}_{\indexbeam}}
\newcommand{\qbeamet}{\vv{d}^{B,t}_{\indexbeam}}

\newcommand{\qxbeamer}{\vv{x}^{B,r}_{\indexbeam}}
\newcommand{\qxbeamet}{\vv{x}^{B,t}_{\indexbeam}}
\newcommand{\qlagrange}{\vv{\lambda}}
\newcommand{\qlagrangee}{\vv{\lambda}_{\indexlagrange}}

\newcommand{\dqsolid}{{\delta \vv{d}^S}}

\newcommand{\dqbeam}{{\delta \vv{d}^B}}

\newcommand{\dqlagrange}{\delta \vv{\lambda}}
\newcommand{\dqlagrangee}{\delta \vv{\lambda}_{\indexlagrange}}
\newcommand{\resbeam}{\vv{r}^B}
\newcommand{\ressolid}{\vv{r}^S}

\newcommand{\gcouplingh}{\Gamma_{c,h}^B}
\newcommand{\gcouplinghsolid}{\Gamma_{c,h}^S}
\newcommand{\mapping}{\chi_h}
\newcommand{\mappingoperation}{\circ \mapping}

\newcommand{\intsolid}[1]{\int_{\osolidO}{ #1 \, \dV}\,}
\newcommand{\intsolidsurface}[1]{\int_{\Gsolids}{ #1 \, \dA}\,}
\newcommand{\intbeamsurface}[1]{\int_{\gcouplingf}{ #1 \,\dA}\,}
\newcommand{\intbeamcenterline}[1]{\int_{\gcouplingl}{ #1 \,\mathrm d \sbeam}\,}
\newcommand{\intbeamcenterlineDomain}[1]{\int_{\obeamL}{ #1 \,\mathrm d \sbeam}\,}

\newcommand{\intbeamcenterlinehe}[1]{\int_{\gcouplingh}{ #1 \,\mathrm d \sbeam}\,}

\newcommand{\beamEint}{\Pi_{\mathrm{int}\,,\placeholder}}
\newcommand{\beamBodyLoad}{\tilde{\tns{f}}}

\newcommand{\srEint}{\Pi_{\mathrm{int}\,,\mathrm{SR}}}
\newcommand{\srTension}{\tns{\Gamma}}
\newcommand{\srTensionC}{\tnss{C}_F}
\newcommand{\srBending}{\tns{\Omega}}
\newcommand{\srBendingC}{\tnss{C}_M}

\newcommand{\klEint}{\Pi_{\mathrm{int}\,,\mathrm{KL}}}
\newcommand{\klTension}{\epsilon}
\newcommand{\klBending}{\tns{\Omega}}
\newcommand{\klBendingC}{\tnss{C}_M}

\newcommand{\tfEint}{\Pi_{\mathrm{int}\,,\mathrm{TF}}}
\newcommand{\tfTension}{\epsilon}
\newcommand{\tfBending}{\kappa}

\newcommand{\Nsolid}{N_\indexsolid}
\newcommand{\Nbeam}{\mat{H}_\indexbeam}
\newcommand{\Nbeamr}{H_\indexbeam^r}
\newcommand{\Nbeamt}{H_\indexbeam^t}
\newcommand{\Nlagrange}{\Phi_\indexlagrange}

\newcommand{\xisolid}{\xi^S}
\newcommand{\etasolid}{\eta^S}
\newcommand{\zetasolid}{\zeta^S}
\newcommand{\xibeam}{\xi^B}

\newcommand{\D}{\mat{D}}
\newcommand{\Dlocal}{\mat{D}^{(\indexlagrange, \indexbeam)}}
\newcommand{\M}{\mat{M}}
\newcommand{\Mlocal}{\mat{M}^{(\indexlagrange, \indexsolid)}}

\renewcommand{\fint}{\vv{f}_{\mathrm{int}}}
\newcommand{\fintsolid}{\fint^S}
\newcommand{\fintbeam}{\fint^B}
\newcommand{\fext}{\vv{f}_{\mathrm{ext}}}
\newcommand{\fextsolid}{\fext^S}
\newcommand{\fextbeam}{\fext^B}
\newcommand{\fc}{\vv{f}_{c}}
\newcommand{\fcsolid}{\fc^S}
\newcommand{\fcbeam}{\fc^B}
\newcommand{\gmt}{\vv{g}_{c}}

\newcommand{\pen}{\epsilon}
\newcommand{\Kss}{\mat{K}_{SS}}
\newcommand{\Kbb}{\mat{K}_{BB}}
\newcommand{\scaling}{\mat{\kappa}}
\newcommand{\scalinglocal}{\scaling^{(\indexlagrange, \indexlagrange)}}

\newcommand{\error}{\norm{e}_{\L2}}

\newcommand{\bhat}{\hat{\tns{b}}}
\newcommand{\that}{\hat{\tns{t}}}

\newcommand{\load}{\hat{t}}
\newcommand{\OmegaSolid}{\Omega^S}
\newcommand{\OmegaBeam}[1]{\Omega^{B#1}}
\newcommand{\uBeam}[1]{\tns{u}_{B#1}}
\newcommand{\uSolid}{\tns{u}_{S}}



%% file: tex/abstract.tex
\begin{abstract}
In this work we present a novel computational method for embedding arbitrary curved one-dimensional (1D) fibers into three-dimensional (3D) solid volumes, as \eg in fiber-reinforced materials.
The fibers are explicitly modeled with highly efficient 1D geometrically exact beam finite elements, based on various types of geometrically nonlinear beam theories.
The surrounding solid volume is modeled with 3D continuum (solid) elements.
An embedded mortar-type approach is employed to enforce the kinematic coupling constraints between the beam elements and solid elements on non-matching meshes.
This allows for very flexible mesh generation and simple material modeling procedures in the solid, since it can be discretized without having to capture for the reinforcements, while still being able to account for complex nonlinear effects due to the embedded fibers.
Several numerical examples demonstrate the consistency, robustness and accuracy of the proposed method, as well as its applicability to rather complex fiber-reinforced structures of practical relevance.
\keywords{Beam-to-solid / 1D-3D coupling \and Finite element method \and Nonlinear beam theory \and Mortar methods}
\end{abstract}

%% file: tex/introduction.tex
\section{Introduction}

Embedding fiber reinforcements into a solid matrix material is a commonly used approach to improve the mechanical behavior of engineering structures.
In many cases, the reinforcements can be considered as being one-dimensional (1D), \ie one dimension is much larger than the other two.
Applications can be found in different fields, such as civil engineering, where steel reinforcements are embedded into concrete to improve its low tensile strength.
In mechanical engineering, fiber-reinforced composites take advantage of fibers with high stiffness by embedding them inside a softer matrix material.
This results in lightweight structures that are used in various applications, such as spacecrafts, boats, or sports equipment.
Last but not least, also nature exploits the benefits of fiber-reinforced materials, as can be seen for example in arterial wall tissue with collagen fibers.
Numerical simulation of such engineering and biomechanical structures is of high importance during the development and design phase, but it is also quite challenging.

\begin{figure}
	\newcommand{\textpos}{(0.6cm,0.1cm)}
	\centering
	\subfigure[]{\label{fig:int_model_a}\includegraphics[resolution=300]{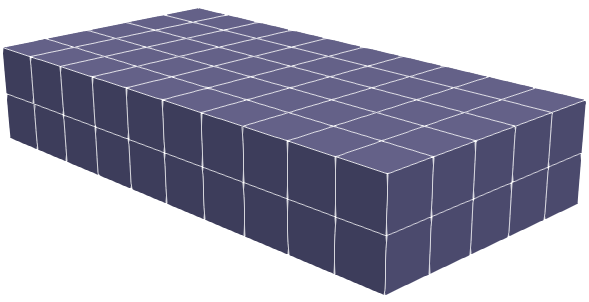}}
	\hfill
	\subfigure[]{\label{fig:int_model_b}\includegraphics[resolution=300]{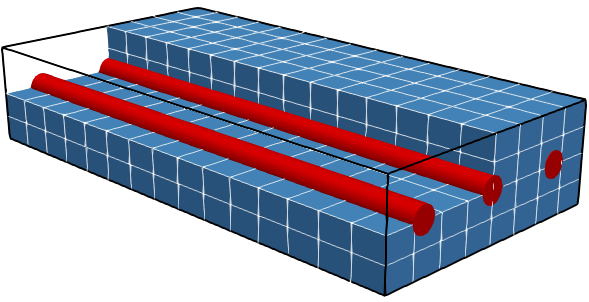}}
	\hfill
	\subfigure[]{\label{fig:int_model_c}\includegraphics[resolution=300]{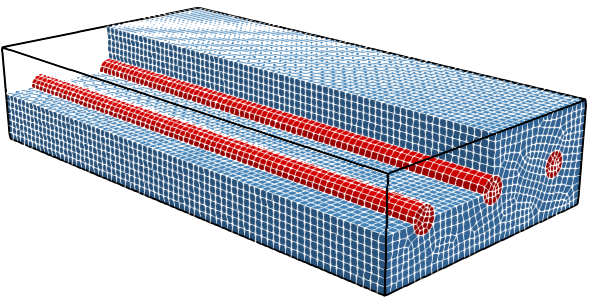}}
	\caption{
		Illustration of different \bis modeling techniques for the same physical problem of a material matrix with embedded fibers.
		Homogenized 3D model \subref{fig:int_model_a}, 1D beams overlapping with 3D volume \subref{fig:int_model_b} and full 3D model \subref{fig:int_model_c}.
		The modeling complexity increases from left to right.
	}
	\label{fig:int_model}
\end{figure}

Different modeling techniques exist to create a numerical model of the reinforced materials, almost all of them being based on the finite element method.
From a mechanical point of view the matrix surrounding the beams is a three-dimensional (3D) continuum, which we will refer to as solid.
In this work, we will denote the combined problem of arbitrary curved beams being embedded inside the solid volume as a \emph{\btsvc} problem.
Figure~\ref{fig:int_model} illustrates different \bis modeling techniques on the basis of the same physical problem of three fibers being embedded inside a material matrix, with the modeling complexity increasing from left to right.
In the model shown in Figure~\ref{fig:int_model_a}, the stiffness contributions from the fibers and matrix are homogenized, thus resulting in an anisotropic material law for the combined volume \cite{Agarwal2017, Wiedemann2007}.
In this case, the fibers are not explicitly modeled, and therefore, this is the most simple case of the models shown in Figure~\ref{fig:int_model} regarding modeling effort and computational complexity.
The main complexity in this approach lies in the accurate homogenization of the fibers and the matrix material.

Figure~\ref{fig:int_model_c} shows a model with the fiber volume explicitly cut out of the solid volume.
This yields a fully 3D surface-to-surface mesh tying problem between the fiber surfaces and the corresponding solid surfaces.
In the shown model the coupling is realized by discretizing the solid and fibers with matching meshes.
Alternatively, the interfaces could also be tied together with coupling methods for non-matching meshes, \eg mortar finite element methods \cite{Puso2004, Puso2008, Popp2009, Popp2012a}.
Even in the non-matching case, the creation of the finite element mesh with explicit boundaries at the interface between beam surface and solid can be a non-trivial task.
The extended finite element method (\xfem) \cite{Moes2003} or immersed finite element methods \cite{Leichner2019b, Rueberg2016} have been used to overcome this issue by implicitly defining the interface between beam surface and solid.
Therefore, a very simple, in many cases even structured Cartesian finite element mesh can be employed.
The drawback of those approaches are the numerically expensive cutting procedures required to implicitly model the interface.
An approach as in Figure~\ref{fig:int_model_c} is the closest to the real physical \btsvc problem and is expected to provide very accurate solutions also close to the interface between fiber and matrix.
Yet, it results in a complex model and an expensive numerical simulation, since resolving the fibers as 3D continua increases the system size by several orders of magnitude.
This limits the applicability for large-scale engineering structures.
Figure~\ref{fig:int_model_b} shows a model with explicitly modeled fibers embedded into the matrix.
In this case, the 1D reinforcements are modeled with a beam theory, which provides accurate and efficient numerical models for the fibers \cite{Reissner1972, Meier2019, Simo1986a, Simo1986b, Meier2017, Meier2014}.
All kinematic fields of the beams are defined along the 1D centerline of the beam.
The solid is modeled, and in particular it is meshed, without subtracting the beam volume from the solid volume, thus resulting in overlapping volumes.
This introduces a modeling error, since in the physical problem no two material points can share the same spatial position.
This modeling error is proportional to the fiber volume fraction as well as the stiffness ratio of fiber and matrix.
The high fiber stiffness compared to the matrix stiffness in the considered cases reduces the influence of this modeling error.
The new \btsvc approach we present in this work follows the modeling ideas from Figure~\ref{fig:int_model_b} and will exclusively use 1D beam formulations to model the fibers.
The resulting \bis model boils down to a mixed-dimensional 1D-3D coupling problem.
Early work on 1D-3D coupling of structures has been carried out in the context of reinforced concrete in \cite{Phillips1976}, with the restriction that the reinforcements have to align with a parameter coordinate of the solid element.
In \cite{Chang1987}, this approach was extended to straight reinforcements with arbitrary directions relative to the solid elements, and in \cite{Elwi1989, Ranjbaran1996, Gomes2001} also curved reinforcements are considered.
All of those mentioned previous works do not introduce additional degrees of freedom for the reinforcements, but instead incorporate the beam stiffness contributions into the stiffness matrices of the solid elements.
Alternatively, the beam degrees of freedom can be kept in the discrete system, which introduces the need for kinematic coupling constraints acting on the beams and solid \cite{Yip2005, Barzegar1997, Kang2014, Durville2007, Kerfriden2019}.
A collocation method is used in \cite{Durville2007} to couple 1D beams into a 3D matrix.
In \cite{Kerfriden2019}, a CutFEM approach is employed to embed 1D structural elements without bending stiffness into a 3D matrix material.
The application of 1D-3D coupling can also be found in other fields than solid mechanics \cite{DAngelo2008, Kremheller2019, Koeppl2018}.
For example, vascular tumor growth is simulated in \cite{Kremheller2019} by coupling the 1D vasculature to the surrounding 3D tissue.
The approach recently presented in \cite{Le2017} combines the techniques from Figure~\ref{fig:int_model_b} and \ref{fig:int_model_c} by using a 3D representation of the beams in zones of interest and 1D structural models otherwise.

\begin{figure}
	\centering
	\subfigure[]{\includegraphics[scale=1]{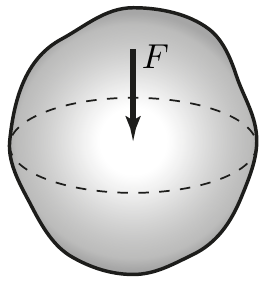}\label{fig:int_kelvin_3D}}
	\hfil
	\subfigure[]{\includegraphics[scale=1]{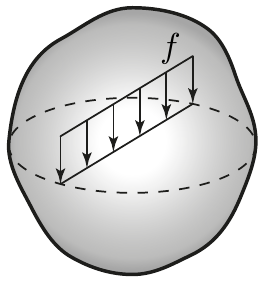}\label{fig:int_kelvin_2D}}
	\caption{
		The 3D Kelvin problem of a force acting on an infinite solid \subref{fig:int_kelvin_3D} and the plane 2D Kelvin problem of a line load acting on an infinite solid \subref{fig:int_kelvin_2D}.
	}
	\label{fig:int_kelvin}
\end{figure}

In this work, we use \C1-continuous geometrically exact beam finite elements \cite{Meier2019}.
Moreover, we propose an embedded 1D-3D mortar-type approach to model the coupling interaction between beam and solid finite elements.
Specifically, a Lagrange multiplier field, representing a line load, is defined along the beam centerline to enforce the coupling constraints, similar to \cite{Kremheller2019}.
The coupling constraints are therefore formulated in a weak variational sense.
The definition of interaction forces between beam and solid as a line load is a problem similar to the plane Kelvin problem of a line load acting on an infinite solid \cite{Kelvin1848, Podio-Guidugli2014, Favata2012}, which is illustrated in Figure~\ref{fig:int_kelvin}.
The exact solution to the Kelvin problem contains singularities in the stress and displacement fields close to the point of action of the line load.
This has a major impact on the well-posedness and applicability of the proposed 1D-3D coupling method, and to the best of the author's knowledge, this aspect along with the resulting spatial convergence behavior will be discussed in detail for the first time.
In the range of our modeling assumptions, \ie relatively high beam stiffness compared to the solid stiffness as well as relatively small beam \cs dimensions compared to the solid finite element sizes, the presented \btsvc method is well-posed, yields very accurate results and exhibits optimal spatial convergence.
In comparison to the available modeling techniques for thin fibers being embedded into a background material,
this allows for an extremely efficient and simple model of the solid phase, while still being able to account for complex nonlinear effects due to the embedded fibers represented by 1D beam formulations.


The remainder of this paper is organized as follows:
In Section~\ref{sec:prob}, we derive the weak form of the quasi-static equilibrium equations for the \btsvc problem via the principle of virtual work.
This is done by combining the individual contributions from 3D solid structures, 1D beams and, in particular, the coupling/interaction terms between them.
In Section \ref{sec:discret}, the finite element method is used to spatially discretize the weak form of the equilibrium equations.
Further, the choice of suitable Lagrange multiplier basis functions as well as numerical integration techniques are discussed.
The final discrete linearized system of equations is then derived by enforcing the coupling constraints in a weighted node-wise manner and by introducing a penalty regularization of the mortar method.
Finally, numerical examples are given in Section~\ref{sec:ex}.
The examples are designed to assess the impact of modeling choices on the quality of the results, as well as to show the applicability of the presented methods to real-life engineering applications. 

%% file: tex/problem_definition.tex
\section{Problem formulation}
\label{sec:prob}

\begin{figure*}
	\centering
	\includegraphics[scale=1]{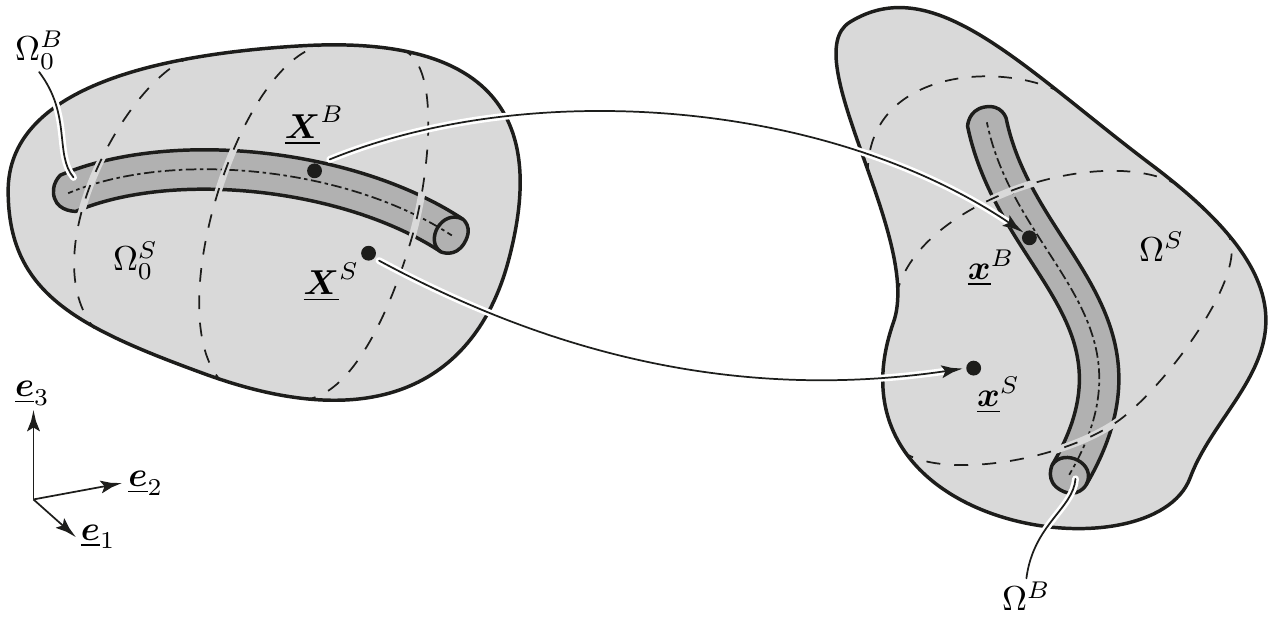}
	\caption{Notation of the finite deformation \btsvc problem.}
	\label{fig:prob_problem}
\end{figure*}

We consider a 3D finite deformation \btsvc problem as shown in Figure~\ref{fig:prob_problem}.
For both the beam and the solid, a Cartesian frame $\{\ex, \ey, \ey\}$ is employed as a fixed frame of reference.
The principle of virtual work (PVW) serves as basis for the employed finite element method.
Contributions to the total virtual work of the system can be split into solid, beam and coupling terms, where the solid and beam terms are independent of the coupling constraint.
Therefore, well-established formulations for the solid as well as the beam can be used without modifications.
Without loss of generality, only quasi-static problems are considered in this work.
This only impacts the virtual work contributions from the solid and the beam, but the coupling terms for the \btsvc problem hold also for time-dependent problems.
The fundamentals of both formulations as well as their individual contribution to the virtual work will be outlined in the next two sections.
Finally the coupling between beam and solid will be described in detail in Section~\ref{sec:prob_coupling}.

\subsection{Solid formulation}
\label{sec:prob_solid}

The solid is modeled as a 3D continuum, represented by the open set $\osolidO \subset \R3$ in the reference configuration and by $\osolidt \subset \R3$ in the deformed configuration.
The reference surface $\osolidOp$ can be divided into the Dirichlet and Neumann boundary surfaces, $\Gsolidu$ and $\Gsolids$, respectively.
In the current configuration they are denoted as $\gsolidu$ and $\gsolids$.
In the reference configuration, a material point on the solid can be identified by its reference position $\Xsolid$.
The current position $\xsolid$ is related to the reference position through the displacement field $\usolid$ via
\begin{equation}
\xsolid\br{\Xsolid} = \Xsolid + \usolid\br{\Xsolid}.
\end{equation}
The variational formulation of the quasi-static balance equations serves as basis for the finite element method, resulting in the solid contribution $\dWs$ to the total virtual work.
A Lagrangian formulation is used, \ie all field variables refer to the reference configuration.
Hence, the integration of the field variables is performed over the reference volume $\osolidO$ and its boundary $\osolidOp$.
Since the variation along the Dirichlet boundary $\Gsolidu$ vanishes, the only remaining surface integral in the variational formulation is over the Neumann boundary $\Gsolids$.
The virtual work $\dWs$ of the solid is given by
\begin{equation}
\dWs
=
\underbrace{
\intsolid{\Spk : \dE}
}_{-\dWints}
\underbrace{
-
\intsolid{\bhat \cdot \dusolid}
-
\intsolidsurface{\that \cdot \dusolid}
}_{-\dWexts}
,
\end{equation}
where $\delta$ denotes the variation of a quantity, $\Spk$ the second Piola-Kirchhoff stress tensor and $\E$ the energy-conjugate Green-Lagrange strain tensor.
Contributions to the external virtual work $\dWexts$ result from the prescribed body load $\bhat$ and surface traction $\that$, both in defined the reference configuration.
The Green-Lagrange strain tensor $\E$ is given as
\begin{equation}
\E = \frac{1}{2}\br{\F\tr \F - \tnssI},
\end{equation}
with $\F = \frac{\partial \xsolid}{\partial \Xsolid}$ being the material deformation gradient and $\tnssI$ the 3D second-order identity tensor.
For simplicity, we assume a hyperelastic material with the strain energy function $\Psi(\E)$, which relates to the second Piola-Kirchhoff stress tensor as follows:
\begin{equation}
\Spk
=
\frac{\partial \Psi(\E)}{\partial \E}.
\end{equation}
All the subsequent examples in Section~\ref{sec:ex} employ a hyperelastic material model for the solid, but this is by no means a requirement of \btsvc, which for example can also be used for elasto-plastic solids.

\subsection{Beam formulation}
\label{sec:prob_beam}

\begin{figure*}
	\centering
	\includegraphics[scale=1]{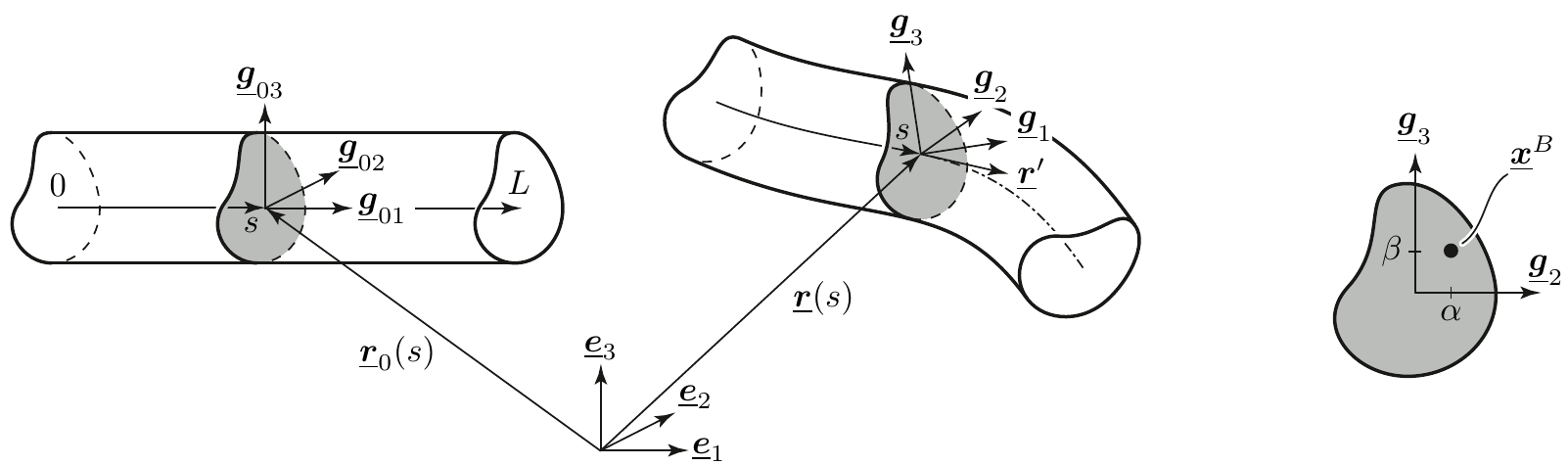}
	\caption{Kinematics of a geometrically exact beam.}
	\label{fig:prob_beam}
\end{figure*}

The beams used in this work are based on the geometrically exact beam theory, which in turn builds upon the kinematic assumption of plane, rigid \cs{s}.
Figure~\ref{fig:prob_beam} shows the reference and current configuration of the beam without any additional kinematic assumptions.
For illustration purposes, the reference configuration shows a straight beam, but unless stated otherwise, the presented beam theories can also be applied to beams with initial curvature.
The complete beam kinematics can be defined by a centerline curve $\rbeam(\sbeam) \in \R3$, connecting the \cs centroids, and a field of right-handed orthonormal triads $\triad(\sbeam):=(\g{1}(\sbeam), \g{2}(\sbeam), \g{3}(\sbeam)) \in \SO$ defining the rotation of the \cs{s}.
Here $s \in [0, \beamlength] =: \obeamLO \subset \R{}$ is the arc-length along the undeformed beam centerline and $\triad(\sbeam)$ is a rotation tensor, which maps the global Cartesian basis vectors $(\ex, \ey, \ez)$ onto the local \cs basis vectors $(\g{1}(\sbeam), \g{2}(\sbeam), \g{3}(\sbeam))$.
The kinematic quantities $\Xbeam, \xbeam, \ubeam\in \R3$, \ie reference position, current position and displacement of an arbitrary point within the \cs, are functions of the centerline coordinate $\sbeam$ as well as the \cs coordinates $\csa, \csb \in \R{}$:
\begin{align}
\Xbeam(\sbeam, \csa, \csb) &= \rbeamO(\sbeam) + \csa \g{02}(\sbeam) + \csb \g{03}(\sbeam),
\\
\xbeam(\sbeam, \csa, \csb) &= \rbeam(\sbeam) + \csa \g{2}(\sbeam) + \csb \g{3}(\sbeam),
\\
\ubeam(\sbeam, \csa, \csb) &= \ubeamr(s) + \csa \br{\g{2}(\sbeam) - \g{02}(\sbeam)} + \csb \br{\g{3}(\sbeam) - \g{03}(\sbeam)} ,
\end{align}
where $\ubeamr = \rbeam - \rbeamO$ is the displacement of the beam centerline.

In this work, three different geometrically exact beam theories are employed.
Their basic kinematic assumptions as well as the corresponding internal elastic energy $\beamEint$ will be stated in the following subsections.
The beam contribution to the global virtual work reads
\begin{equation}
\dWb_{\placeholder} =
\underbrace{
	\delta \beamEint
}_{-\dWb_{\mathrm{int}}}
\underbrace{
	-\intbeamcenterlineDomain{\delta \rbeam \cdot \beamBodyLoad}
	-\dWb_{\mathrm{ext}\,,\placeholder}
}_{-\dWb_{\mathrm{ext}}},
\label{eq:prob_dWb}
\end{equation}
where the term $\intbeamcenterlineDomain{\delta \rbeam \cdot \beamBodyLoad}$ is the virtual work of distributed line loads $\beamBodyLoad$ along the beam and is independent of the specific beam formulation, because it only depends on centerline degrees of freedom.
The virtual work of external forces and moments at the Neumann boundaries as well as of distributed moments is summarized in $\dWb_{ext,\placeholder}$.
These contributions depend on the rotational field along the centerline and therefore also differ for the three employed beam theories.
A consistent and objective handling of the rotational variations contained in $\dWb_{ext,\placeholder}$ is a non-trivial task.
Since it is not the main aspect of the current work, the interested reader is referred to \cite{Meier2019}.
To improve readability of the following equations, a derivative with respect to the beam centerline coordinate $s$ will be represented by $\placeholder' := \partial \placeholder / \partial \sbeam$ throughout this section.

\subsubsection{\sr beam theory}

Of the three beam theories considered in this work, the \sr (SR) beam theory is the most general one, as it does not introduce additional kinematic constraints on the beam.
This results in shear-deformable beams capturing six modes of deformation: axial strain, two bending modes, torsion and two shear modes.
The \cs kinematics can be described with six degrees of freedom: the spatial position of the \cs $\rbeam(\sbeam)$ and its rotation vector $\rotvec(\sbeam) \in \R3$, which defines the \cs triad $\triad(\sbeam) = \triad(\rotvec(\sbeam))$ based on the well-known Rodrigues formula \cite{Meier2019}.
The internal elastic energy of the beam is given as
\begin{equation}
\srEint = \frac{1}{2} \intbeamcenterlineDomain{
	\srTension\tr \srTensionC \srTension
	+
	\srBending\tr \srBendingC \srBending
	},
\end{equation}
where axial tension and shear strains are represented by the material deformation measure $\srTension := \srTension(\rotvec, \rbeam') = \triad\tr \rbeam' - \ex \in \R3$, while torsion and bending are represented by the material curvature vector $\srBending \in \R3$, which in turn follows from $\srBending \times \tns{a} = \triad\tr \triad' \tns{a}\ \forall\ \tns{a} \in \R3$.
Using the rotation vector parameterization of the triad field $\triad(\rotvec(\sbeam))$ as discussed above, the resulting curvature vector can be formulated as a function of $\rotvec$ and $\rotvec'$, \ie $\srBending = \srBending(\rotvec, \rotvec')$.
The constitutive matrices $\srTensionC$ and $\srBendingC$ are defined as
\begin{equation}
\srTensionC = \begin{bmatrix}EA & & \\ & GA_2 & \\ & & GA_3\end{bmatrix}
\quad
\text{and}
\quad
\srBendingC = \begin{bmatrix}GI_T & & \\ & EI_2 & \\ & & EI_3\end{bmatrix},
\end{equation}
where $E$ is the Young's modulus, $G$ the shear modulus, $A$ the \cs area, $A_2$ and $A_3$ the effective shear areas, and $I_T, I_2, I_3$ are the polar and planar second moments of area, respectively.

\subsubsection{\kl beam theory}
The \kl (KL) theory introduces an additional kinematic constraint, restraining the shear deformation of the beam.
This is equivalent to the requirement that the first \cs basis vector $\g1$ is parallel to the centerline tangent $\rbeam'$, or
\begin{equation}
\g2 \cdot \rbeam' \equiv 0
\quad
\wedge
\quad
\g3 \cdot \rbeam' \equiv 0.
\end{equation}
While the position of the \cs is described in the same manner as for the \sr beam, the additional constraints reduce the number of independent rotations to one, thus a total of four degrees of freedom remain to fully describe the \cs.
The sole remaining rotational degree of freedom $\twist(\sbeam) \in \R{}$ describes the twist rotating of the \cs around the tangent vector $\rbeam'$ measured with respect to a properly defined reference triad $\triad_{\text{ref}}(\rbeam')$, such that the \cs triad can be described as a function of the centerline tangent and the twist, $\triad(s) = \triad(\rbeam'(s), \twist(s))$.
A detailed overview, how to parametrize the twist degree of freedom, can be found in \cite{Meier2019}.
The curvature of the beam centerline is described with the Frenet--Serret vector
\begin{equation}
\frenet = \frac{\rbeam' \times \rbeam''}{\norm{\rbeam'}^2},
\end{equation}
which only depends on the beam centerline.
The definition of the curvature contains second derivatives of the beam centerline, therefore resulting in the smoothness requirement of $\C1$ continuous centerlines.
Defining the material curvature vector $\klBending$ identical to the \sr case above, it can be formulated as a function of $\twist$, $\twist'$, $\rbeam'$ and $\frenet$ for the \kl case \cite{Meier2019}, \ie $\klBending = \klBending(\twist, \twist', \rbeam', \frenet)$.
The internal energy for the \kl beam is
\begin{equation}
\klEint = \frac{1}{2} \intbeamcenterlineDomain{
	EA \klTension^2\
	+
	\klBending\tr \klBendingC \klBending
}.
\end{equation}
Therein, $\epsilon = \norm{\rbeam'} - 1$ is the axial tension of the beam.

\subsubsection{\Tf beam theory}
The \tf (TF) beam formulation considered in this work was first proposed in \cite{Meier2015} and extended in \cite{Meier2016a}.
It represents a special case of the Kirchhoff-Love beam theory.
For certain properties of the problem, \ie straight undeformed beams with axisymmetric \cs[s] and no external torsional moments, it can be shown that the static equilibrium configurations resulting from the \kl beam theory are characterized by (exactly) vanishing torsion \cite{Meier2015}.
The fact that these requirements are fulfilled in many practically relevant systems, and also in most of the examples considered in this work, justifies and motivates the application of this type of beam element formulation.
Compared to the \kl beam, the twist degree of freedom is not present anymore and the beam can be completely described by its centerline position, \ie three degrees of freedom per \cs.
Since the discrete representation and algorithmic treatment of large rotations is the main complexity of geometrically nonlinear beam theories, the employed torsion-free beam theory, which can completely abstain from any rotational degrees of freedom, is particularly appealing and easy to handle.
The internal energy of the \tf beam reads
\begin{equation}
\tfEint = \frac{1}{2} \intbeamcenterlineDomain{
	EA \tfTension^2\
	+
	EI \tfBending^2
},
\end{equation}
with the scalar curvature $\kappa = \norm{\frenet}$.

\subsection{\Btsvc}
\label{sec:prob_coupling}

In the \btsvc problem shown in Figure~\ref{fig:prob_problem}, the beam is embedded inside the solid volume.
The most natural choice for the coupling conditions is to couple the beam surface $\obeamOp$ to the solid volume $\osolidO$.
However, there is no explicit surface in the solid domain, to define the coupling conditions on.
Therefore, this is a surface-to-volume (2D-3D) coupling problem, \ie the beam surface is embedded into the background solid volume.
The coupling constraints are formulated in the reference configuration and read
\begin{equation}
\label{eq:prob_constraint_strong}
\ubeam - \usolid = \tnsO \quad \text{on} \quad \gcouplingf,
\end{equation}
with $\gcouplingf = \obeamOp$ being the coupling surface.
The Lagrange multiplier method is employed to impose the coupling constraint.
A Lagrange multiplier vector field $\lagrangef(\sbeam, \csa, \csb) \in \R3$ is defined on $\gcouplingf$, which can be interpreted as the negative interface tractions acting on the beam surface.
Contributions to the total virtual work are the coupling interface contribution
\begin{equation}
\label{eq:dWmt}
- \dWmtf
=
\intbeamsurface{ \lagrangef \left( \dubeam - \dusolid \right) },
\end{equation}
and the variational form of the coupling constraints
\begin{equation}
\label{eq:dWlambda}
\dWlambdaf
=
\intbeamsurface{ \dlagrangef \left( \ubeam - \usolid \right) }.
\end{equation}
This leads to a saddle point-type weak formulation of the 2D-3D \btsvc problem:
\begin{equation}
\label{eq:pvwf}
\dWs + \dWb - \dWmtf + \dWlambdaf = 0.
\end{equation}

The integrals in equations \eqref{eq:dWmt} and \eqref{eq:dWlambda} are evaluated on the coupling surface $\gcouplingf$, which requires a computationally expensive numerical integration of $\dWmtf$ and $\dWlambdaf$.
For the inherent assumption in this work, that the \cs dimensions of the beam are small compared to the other dimensions of the \btsvc problem, we can approximate the surface integrals as line integrals along the beam axis $\obeamLO$.
These line integrals can be evaluated very efficiently.
The approximation changes the physical coupling dimensionality applied to the \btsvc model from surface-to-volume (2D-3D) to a line-to-volume (1D-3D) coupling.
The new coupling domain is $\gcouplingl = \obeamLO$.
Since this is a significant change in the mathematical description of the mechanical model, the implications of this choice will be discussed in several remarks at the end of this section.
The approximated variational coupling terms read
\begin{align}
\label{eq:dWmtl}
\dWmtf
&\approx
\dWmtl
=
\intbeamcenterline{\lagrangel \left( \dubeamr - \dusolid \right)},
\\
\label{eq:dWlambdal}
\dWlambdaf
&\approx
\dWlambdal
=
\intbeamcenterline{\dlagrangel \left( \ubeamr - \usolid \right)}.
\end{align}
Here $\lagrangel(\sbeam) \in \R3$ is a new Lagrange multiplier field defined along the beam centerline.
We point out that $\lagrangef$ and $\lagrangel$ have different physical dimensionality and, accordingly, also different units: the first one is a surface load, while the latter one represents a line load.
The final PVW for the 1D-3D \btsvc problem reads
\begin{equation}
\label{eq:problem_pvwl}
\dWs + \dWb - \dWmtl + \dWlambdal = 0.
\end{equation}
For improved readability, the superscript 1D-3D for the line-to-volume coupling terms will be omitted from now on.

\begin{remark}
	In the previous considerations, it was assumed, without loss of generality, that the beam consists of a single fiber which lies completely inside the solid.
	The derived equations also hold if the beam sticks out of the solid volume.
	In this case the coupling integrals are not evaluated on the complete beam domain, but instead only on the portion of the beam centerline inside the solid.
	The impact on the numerical integration will discussed in Section~\ref{sec:discret_num_integration}.
\end{remark}

\begin{remark}
	With the definition of the line-to-volume coupling terms in \eqref{eq:dWmtl} and \eqref{eq:dWlambdal}, the coupling is now exclusively formulated through the beam centerline position, which decouples the \cs rotations $\triad$ from the solid deformations.
	In particular, relative rotations between the \cs and the solid around the tangent vector $\rbeam'$ are not restrained.
	At first glance, this might be considered as a rather coarse approximation for certain physical systems such as fiber-reinforced composite materials, where fibers are \eg molded / glued into a matrix such that all modes of relative motion are blocked.
	However, in our target applications, the main contributions to the internal energy of the beams and the mechanical resistance of the overall structure stem from bending and axial tension of the fibers, therefore justifying the choice to neglect the coupling of \cs rotations.
	Additionally, this will lead to coupling terms, which only contain the centerline degrees of freedom and are independent of the actual beam theory.
	This allows for an easy adaptation of our \btsvc method to different beam theories.

	As a further consequence, the rotation of beam fibers around their centerline might be unconstrained, possibly yielding a singular linear system to solve.
	In practice, this repairable deficiency is limited to static analyses and the undeformed configuration.
	As a remedy, one either imposes Dirichlet boundary conditions on at least one of the twist degrees of freedom	or uses standard linear solvers with deflation capabilities to properly exclude such nullspace modes.	
	The problem is cured as soon as the beam centerlines have deformed, \ie usually after the first Newton step.
	This discussion also underlines an advantage coming with the torsion-free beam theory: The corresponding beam finite elements do not have any rotational degrees of freedom, and consequently, such rigid body modes cannot occur.
\end{remark}

\begin{remark}
	\label{rmk:singular_solutions}
	Another aspect to be addressed, when switching from 2D-3D to 1D-3D coupling, is the introduction of singular solutions.
	From a mechanical point of view, the line-to-volume coupling is equivalent to a line load inside the solid.
	This is a generalized version of the Kelvin problem \cite{Podio-Guidugli2014, Kelvin1848, Favata2012}, which consists of an infinite solid loaded with an embedded line load.
	The analytical solution for the Kelvin problem has a singularity at the line load point of action, not only in the stress field, but also in the displacement field.
	This has a significant impact on the spatial convergence of the finite element discretization and will be discussed in detail in Section~\ref{sec:ex_spatial_convergence}.	
\end{remark}

%% file: tex/discretization.tex
\section{Spatial discretization and numerical integration}
\label{sec:discret}

An isoparametric finite element discretization is employed to approximate the continuous fields for geometry, displacement as well as virtual displacement.
The interpolation of the positions and displacements in the solid domain is given by
\begin{equation}
\label{eq:discret_fe_solid}
\Xsolidh = \sumsolid{\Nsolid\br{\xisolid, \etasolid, \zetasolid} \qxsolide}
\end{equation}
and
\begin{equation}
\label{eq:discret_fe_solid_displacement}
\usolidh = \sumsolid{\Nsolid\br{\xisolid, \etasolid, \zetasolid} \qsolide}.
\end{equation}
Here, $\Nsolid \in \R{}$ is the finite element shape function for the solid node $\indexsolid$, $\qxsolide \in \R3$ and $\qsolide \in \R3$ are the nodal reference position and displacement, respectively.
The total number of solid nodes is $\nsolid$.
The variables $\xisolid$, $\etasolid$ and $\zetasolid$ are the 3D coordinates of the solid finite element parameter space.

\kl and \tf beam elements require a $\C1$-continuous centerline interpolation, which is realized with third-order Hermite polynomials \cite{Vetyukov2014, Meier2019}.
An objective and path-independent interpolation of the rotational field $\triad_h^\esolid(\xibeam)$ along the beam centerline is a non-trivial task and will not be discussed here, since the rotations do not appear in the coupling terms anyway.
A comprehensive overview on this topic can be found in \cite{Meier2019}.
The resulting beam element has two centerline nodes with six degrees of freedom per node, \ie three positional and three tangential degrees of freedom.
Due to its superior numerical properties, this discretization scheme is also used for the \sr beam element, as derived in \cite{Meier2018}.
The beam centerline reference position and displacement are interpolated by
\begin{equation}
\label{eq:discret_fe_beam}
\rbeamOh = \sumbeam{\Nbeamr(\xibeam) \qxbeamer + \Nbeamt(\xibeam) \qxbeamet}
\end{equation}
and
\begin{equation}
\label{eq:discret_fe_beam_displacement}
\ubeamrh = \sumbeam{\Nbeamr(\xibeam) \qbeamer + \Nbeamt(\xibeam) \qbeamet},
\end{equation}
where $\Nbeamr \in \R{}$ and $\Nbeamt \in \R{}$ denote the Hermite shape functions for the positional and tangential degrees of freedom for the beam node $\indexbeam$.
Both shape functions are a function of the scalar beam centerline parameter coordinate $\xibeam$.
The discrete vectors $\qxbeamer, \qxbeamet \in \R3$ are the reference position and tangent, respectively.
The discrete degrees of freedom $\qbeamer, \qbeamet \in \R3$ denote the nodal displacements and tangent increments.
The total number of beam centerline nodes is $\nbeam$.
To improve readability of the following derivations, the beam centerline displacement is redefined in the following way
\begin{align}
\Nbeam &= \begin{bmatrix}\Nbeamr \matI^{3 \times 3} & \Nbeamt \matI^{3 \times 3}\end{bmatrix} \in \R{3 \times 6},
\\
\qbeame &=
\begin{bmatrix}
\qbeamer \\ \qbeamet
\end{bmatrix} \in \R6,
\\
\ubeamrh &= \sumbeam{\Nbeam(\xibeam) \qbeame}.
\end{align}
The discrete nodal displacement vector $\qbeame$ now contains all centerline degrees of freedom for the node $\indexbeam$.

Employing a mortar-type coupling approach, the Lagrange multipliers are also approximated with a finite element interpolation \cite{Wohlmuth2000, Belgacem1999, Popp2010}.
The continuous Lagrange multiplier field $\lagrange$ is defined along the beam centerline.
Therefore, the Lagrange multiplier interpolation is defined along the 1D beam elements.
All subsequent integration is performed on the domain $\gcouplingh$, which is the projection of the beam centerline domain $\obeamLO$ onto the beam finite element function space.
In the nomenclature of classical contact mechanics, the beam would be considered the slave side, and the solid the master side.
The approximated Lagrange multiplier field reads
\begin{equation}
\label{eq:discret_fe_lagrange}
\lagrangeh = \sumlagrange{\Nlagrange(\xibeam) \qlagrangee},
\end{equation}
where $\Nlagrange \in \R{}$ is the shape function for the discrete Lagrange multiplier vector $\qlagrangee \in \R3$ at node $\indexlagrange$.
The total number of discrete Lagrange multiplier nodes is $\nlagrange$, which is not necessarily equal to $\nbeam$.
The shape function $\Nlagrange$ is a function of the scalar beam centerline parameter coordinate $\xibeam$.
Note that even though the Lagrange multipliers are defined along the beam centerline domain, the displacement shape functions $\Nbeam$ will not be used to interpolate the Lagrange multiplier field.
An adequate choice of Lagrange multiplier shape functions will be discussed in Section~\ref{sec:discret_lagrange_shapefunctions}.
The nodal discrete unknowns $\qsolide$, $\qbeame$ and $\qlagrangee$ are assembled into the global degrees of freedom vectors $\qsolid$, $\qbeam$ and $\qlagrange$.

Insertion of the finite element approximations \eqref{eq:discret_fe_solid_displacement}, \eqref{eq:discret_fe_beam_displacement} and \eqref{eq:discret_fe_lagrange} into the variational form of the coupling constraints \eqref{eq:dWlambdal} gives
\begin{multline}
\dWlambdah = 
\sumbeam{\sumlagrange{
	\dqlagrangee\tr
	\br{
		\intbeamcenterlinehe{
			\Nlagrange
			\Nbeam
		}
	}
	\qbeame
}}
\\-
\sumsolid{\sumlagrange{
		\dqlagrangee\tr
		\br{
			\intbeamcenterlinehe{
				\Nlagrange
				\br{\Nsolid \mappingoperation}
			}
		}
		\qsolide
	}},
\end{multline}
where $\mapping: \gcouplingh \rightarrow \gcouplinghsolid$ defines a suitable projection from a point on the beam centerline to the corresponding point in the solid volume.
In the previous equation, two local matrices with mass matrix-like structure can be identified:
\begin{align}
\label{eq:discret_D_local}
\Dlocal &=
\intbeamcenterlinehe{
	\Nlagrange
	\Nbeam
}\ \in \R{3\times 6},
\\
\label{eq:discret_M_local}
\Mlocal &=
\intbeamcenterlinehe{
	\Nlagrange
	\br{\Nsolid \mappingoperation}
}
\matI^{3 \times 3} \ \in \R{3\times 3}.
\end{align}
There, $\Dlocal$ describes the coupling between the Lagrange multiplier node $\indexlagrange$ and the beam node $\indexbeam$ and $\Mlocal$ describes the coupling between the Lagrange multiplier node $\indexlagrange$ and the solid node $\indexsolid$.
They can be assembled into global, so called mortar matrices $\D \in \R{3 \nlagrange \times 6 \nbeam}$ and $\M \in \R{3 \nlagrange \times 3 \nsolid}$, which both are rectangular in general.
A similar expression containing $\D$ and $\M$ can also be derived for the virtual work $\dWmth$ of the coupling forces. 
All in all, the coupling contributions to the weak form can now be stated in global matrix form
\begin{align}
-\dWmth &=
\dqbeam \tr \underbrace{ \D\tr \qlagrange }_{\fcbeam(\qlagrange)}
-
\dqsolid \tr \underbrace{ \M \tr \qlagrange }_{\fcsolid(\qlagrange)},
\\
\dWlambdah &= \dqlagrange\tr \D \qbeam - \dqlagrange\tr \M \qsolid
=
\dqlagrange\tr
\underbrace{
	\begin{bmatrix}
	-\M & \D
	\end{bmatrix}
	\begin{bmatrix}
	\qsolid \\ \qbeam
	\end{bmatrix}
}_{\gmt(\qsolid, \qbeam)}.
\end{align}
Here, $\fcsolid$ and $\fcbeam$ are the vectors with the discretized coupling forces acting on the solid and beam degrees of freedom, respectively.
The vector $\gmt$ contains the discretized constraint equations and its entries can be interpreted as the relative displacement between beam centerline and solid weighted with the Lagrange multiplier shape functions.
Inserting all discretized variables into \eqref{eq:problem_pvwl} gives the discrete nonlinear system of equations for the quasi-static \btsvc problem:
\begin{align}
\label{eq:discret_system_start}
\fintsolid(\qsolid) + \fcsolid(\qlagrange) - \fextsolid &= \vvO,
\\
\fintbeam(\qbeam) + \fcbeam(\qlagrange) - \fextbeam &= \vvO,
\\
\label{eq:discret_system_end}
\gmt(\qsolid, \qbeam) &= \vvO.
\end{align}
Here, $\fintsolid$ and $\fintbeam$ are the internal force vectors of the solid and beam, respectively.
The Newton--Raphson algorithm is used to obtain solutions to the system of nonlinear equations.
Therefore, a linearization of equations \eqref{eq:discret_system_start} to \eqref{eq:discret_system_end} with respect to the global unknowns $\qsolid$ and $\qbeam$ has to be derived.
The linearized system of equations with saddle point structure reads:
\begin{multline}
\label{eq:discret_global_system}
\begin{bmatrix}
\Kss & \matO & -\M\tr \\
\matO & \Kbb & \D\tr \\
-\M & \D & \matO
\end{bmatrix}
\begin{bmatrix}\Delta \qsolid \\ \Delta \qbeam \\ \qlagrange
\end{bmatrix}
=
\begin{bmatrix}
-\fintsolid(\qsolid) - \fcsolid(\qlagrange) + \fextsolid \\
-\fintbeam(\qbeam) - \fcbeam(\qlagrange) + \fextbeam \\
\vvO
\end{bmatrix}
\\=
- \begin{bmatrix}
\ressolid \\
\resbeam \\
\vvO
\end{bmatrix},
\end{multline}
where $\Kss = \lininline{\fintsolid(\qsolid)}{\qsolid}$ and $\Kbb = \lininline{\fintbeam(\qbeam)}{\qbeam}$ are the stiffness matrices of the solid and beam, respectively.

\subsection{Lagrange multiplier shape functions}
\label{sec:discret_lagrange_shapefunctions}

The choice of Lagrange multiplier shape functions is important for the mathematical properties of the discretized system, since the discrete Lagrange multiplier bases, \ie shape functions, must fulfill an inf-sup condition with the displacement field \cite{Boffi2013}.
In the context of surface-to-surface contact or mesh tying in solid mechanics, this is a well studied-topic.
However, in the considered \btsvc problem, we employ Hermite polynomials as primary shape functions for the slave side, \ie the beam, which is unusual compared to the standard surface-to-surface case.
Additionally, we deal with an embedded 1D-3D coupling, \ie there is no explicit curve representation in the solid mesh to match the beam centerline, which can lead to stability issues \cite{Sanders2012a}.
The numerical experiments in Section~\ref{sec:ex_spatial_convergence} and \ref{sec:ex_penalty_convergence} carefully evaluate the influence of different Lagrange multiplier bases on the numerical properties of the \btsvc problem.

Since the Lagrange multipliers are defined on the beam centerline, a natural choice in the spirit of the mortar method would be to use the same shape functions as for the beam elements, \ie third-order \C1-continuous Hermite polynomials.
However, for neighboring beam elements with equal length the integral over the Hermite shape functions associated with the tangential degrees of freedom becomes zero.
This can lead to numerical difficulties in the constraint enforcement.
Therefore, in this work, standard Lagrangian shape functions are used to interpolate the Lagrange multiplier field.
Three different types of shape functions will be compared: linear, quadratic and cubic.
In surface-to-surface mortar methods, the use of stable lower order interpolations for the Lagrange multipliers compared to the displacement interpolation order was already successfully explored in \cite{Puso2008, Popp2012a}.

\begin{remark}
	The previous derivations are given for the case, where the constraint equations are fulfilled in a truly weak (variational) sense.
	In Section~\ref{sec:ex}, this mortar-type coupling will be compared to a classical \gptslong (\gpts) coupling approach.
	In the \gpts coupling, the strong form of the constraint equations \eqref{eq:prob_constraint_strong} is fulfilled at each Gauss-point along the beam, \ie a discrete 3D Lagrange multiplier vector $\qlagrangee^{\text{\gpts}}$ is defined at each Gauss-point in the sense of a collocation method.
	However, \gpts coupling can also be interpreted as a special case of the mortar coupling, namely if the Lagrange multiplier field is interpolated as
	\begin{equation}
	\lagrangeh^{\text{\gpts}} = \sumlagrange{\tilde{w}_j \dirac\br{\tilde{\xi}^B_j - \xibeam} \qlagrangee^{\text{\gpts}}}.
	\end{equation}
	Here, $\delta$ is the Dirac delta distribution with the property $\intbeamcenterlinehe{\dirac\br{\alpha-\xibeam}f\br{\xibeam}}=f\br{\alpha}$.
	The position and weight of the $\indexlagrange$-th Gauss-point are denoted with $\tilde{\xi}^B_j$ and $\tilde{w}_j$, respectively.
\end{remark}

\subsection{Enforcement of constraint equations}
The constraint equations \eqref{eq:prob_constraint_strong} are discretized with a mortar coupling approach using Lagrange multipliers, this resulting in a mixed formulation.
However, due to certain drawbacks, \eg an increased system size compared to the uncoupled problem and a saddle point structure, \eqref{eq:discret_global_system} will not be solved directly here to obtain solutions to the \btsvc problem.
Instead, the penalty method is used to obtain approximate solutions of \eqref{eq:discret_global_system}.
This results in a formulation that is purely displacement-based and does not contain any additional variables.
The main idea behind this well-known penalty regularization of the mortar method is to allow a relaxation of the discretized coupling constraints $\gmt = \vvO$ in the form
\begin{equation}
\label{eq:discret_penalty}
\qlagrange = \pen \scaling^{-1}\gmt(\qsolid, \qbeam).
\end{equation}
Herein, the Lagrange multipliers are no longer independent variables, but well-defined functions of the beam and solid displacements.
They can subsequently be removed from the global system of equations.
In \eqref{eq:discret_penalty}, $\pen \in \R{+}$ is the penalty parameter and it is clear that for $\pen \rightarrow \infty$, \eqref{eq:discret_penalty} becomes equivalent to \eqref{eq:discret_system_end}.
The entries in the weighted relative displacement vector $\gmt$ are proportional to the support of the corresponding Lagrange multiplier shape function, \ie they depend on the beam element length.
If unaccounted for, this dependency would result in a violation of the basic patch tests presented in Section~\ref{sec:ex_patch}.
To resolve this problem, the relaxation of the constraints in \eqref{eq:discret_penalty} is additionally multiplied with the inverse of the diagonal nodal scaling matrix $\scaling$, similar to the approach in \cite{Yang2005}.
The local scaling matrix for the Lagrange multiplier node $\indexlagrange$ is defined by
\begin{equation}
\scalinglocal =
\intbeamcenterlinehe{\Nlagrange} \matI^{3 \times 3},
\end{equation}
and is assembled into the global scaling matrix $\scaling$.
With the penalty approach, the coupling forces $\fcsolid$ and $\fcbeam$ can be stated as,
\begin{align}
\fcsolid(\qsolid, \qbeam) &=
\pen \M\tr \scaling^{-1} 	\begin{bmatrix}
-\M & \D
\end{bmatrix}
\begin{bmatrix}
\qsolid \\ \qbeam
\end{bmatrix}
\\
\fcbeam(\qsolid, \qbeam) &=
\pen \D\tr \scaling^{-1} 	\begin{bmatrix}
-\M & \D
\end{bmatrix}
\begin{bmatrix}
\qsolid \\ \qbeam
\end{bmatrix}.
\end{align}
With this the final global system of equations \eqref{eq:discret_global_system} becomes:
\begin{equation}
\label{eq:discret_global_system_penalty}
\begin{bmatrix}
\Kss + \epsilon \M\tr \scaling^{-1} \M & -\epsilon \M\tr \scaling^{-1} \D\\
-\epsilon \D\tr \scaling^{-1} \M & \Kbb + \epsilon \D\tr \scaling^{-1} \D
\end{bmatrix}
\begin{bmatrix}\Delta \qsolid \\ \Delta \qbeam
\end{bmatrix}
=
- \begin{bmatrix}
\ressolid \\ \resbeam
\end{bmatrix}.
\end{equation}
Here, the number of global unknowns is the same as in the uncoupled case.
An additional effect of the penalty-regularized version of the mortar method is the elimination of the saddle point structure in the stiffness matrix.
However, there are some drawbacks of the penalty approach.
The constraint equations are violated by definition, which only can be reduced with higher penalty parameters, but this in turn leads to an ill-conditioned tangential system matrix.
Therefore, it is desirable to choose a penalty parameter that results in a sufficiently accurate solution of the constraint equations, but also limits unwanted numerical effects.
The influence of the penalty parameter in practice will be discussed in detail in Section~\ref{sec:ex_penalty_convergence}.

\subsection{Numerical Integration}
\label{sec:discret_num_integration}
\begin{figure*}
	\centering
	\includegraphics[scale=1]{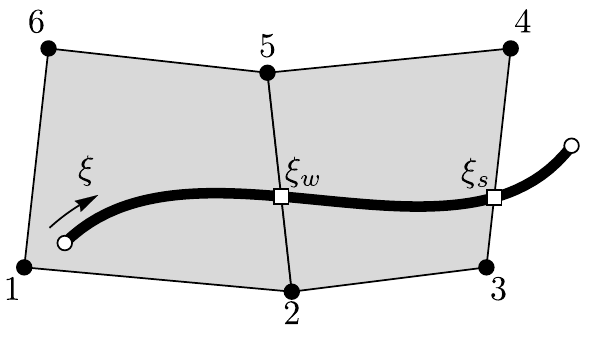}
	\hspace{1cm}
	\includegraphics[scale=1]{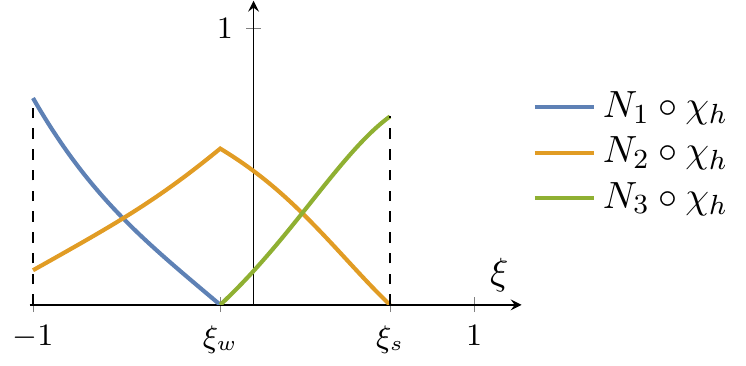}
	\caption{
		Illustration of weak and strong discontinuities.
		Patch of two solid elements and one beam element with a weak discontinuity at $\xi_w$ and a strong discontinuity at $\xi_s$ (left) and the projection of selected solid shape functions onto the beam centerline (right).
	}
	\label{fig:discret_discontinuities}
\end{figure*}

\begin{figure*}
	\centering
	\includegraphics[scale=1]{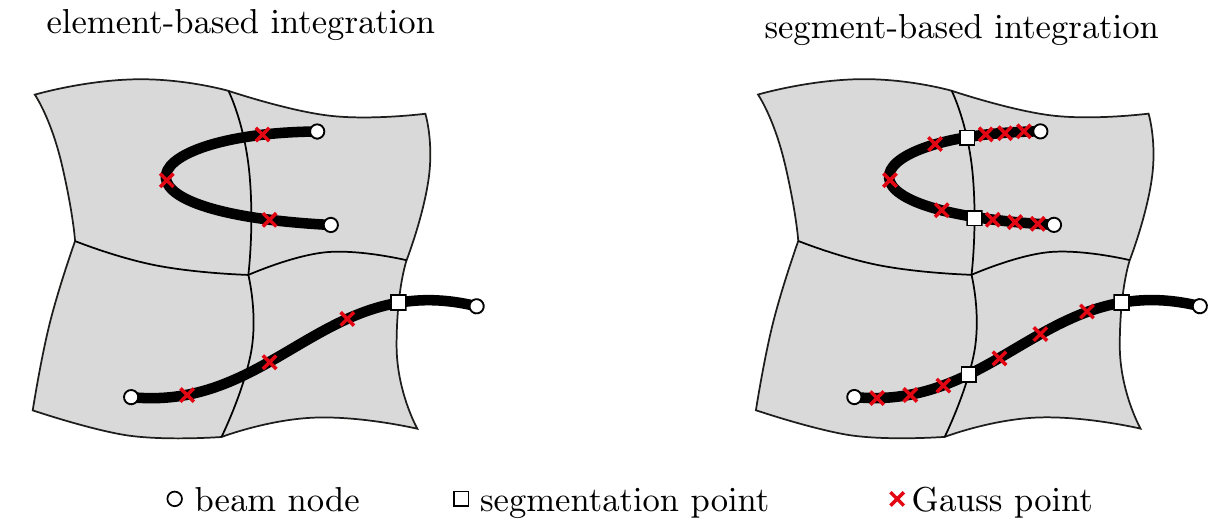}
	\caption{
		Illustration of element-based and segment-based integration.
	}
	\label{fig:discret_segmentation}
\end{figure*}

The \btsvc contributions to the global system of equations are all calculated via integration over the beam domain in the reference configuration, \cf \eqref{eq:discret_D_local} and \eqref{eq:discret_M_local}.
Numerical integration, namely a Gauss--Legendre quadrature, is used to evaluate the coupling matrices $\D$ and $\M$ and the scaling matrix $\scaling$ during the finite element simulation.
An accurate numerical evaluation of the coupling integrals is absolutely essential to pass basic consistency tests, such as the patch tests in Section~\ref{sec:ex_patch}.
The integrands in $\D$ and $\scaling$ solely contain fields defined along the beam centerline, namely the beam displacements and the Lagrange multipliers.
If the Jacobian $\jacobian$ along the beam element is constant, the integrand is of polynomial form and the numerical integration is exact, if enough quadrature points are used.
In the cases considered in this work, the maximal polynomial degree of the integrand in $\D$ and $\scaling$ is 6, \ie third-order beam shape functions and third-order Lagrange multiplier shape functions.
Therefore, 4 Gauss--Legendre points are needed for the numerical integration to be exact.
The integrand of $\M$ contains fields defined along the beam centerline as well as the solid volume.
In Figure~\ref{fig:discret_discontinuities}, it can be seen that the evaluation of the solid shape functions along the beam centerline results in a general nonlinear function which contains so-called weak discontinuities, \ie kinks at the points where the beam crosses between solid elements, and strong discontinuities, \ie jumps at points where the beam sticks out of the solid volume.
Moreover, the continuous parts of the integrand in $\M$ are not of polynomial degree.
To still guarantee high accuracy of numerical integration for the integrand in $\M$, two different algorithms will be investigated and compared, \cf Figure~\ref{fig:discret_segmentation}.
They are illustrated in Figure~\ref{fig:discret_segmentation}.
Element-based integration uses a fixed number of Gauss-points per beam element.
The only exception occurs at strong discontinuities, where the integration is only performed for the part of the beam element inside the solid volume.
In segment-based integration, the integration domain along the beam element is split into multiple segments, such that the integrand in the individual segments does not contain any kinks.
Each segment is then integrated with a fixed number of Gauss-points.

The global coupling matrices only depend on the initial configuration of the \btsvc problem, \ie they remain constant over the course of the simulation.
From a computational point of view, it makes sense to evaluate the coupling matrices $\D$, $\M$ and $\scaling$ once and store them for subsequent Newton iterations and time steps.
Nevertheless, it is important to address the impact of the different numerical integration schemes with regard to computational performance and accuracy.
Independent of the integration scheme used, each Gauss-point evaluation requires the solution of a local nonlinear system of equations, \ie the projection of the point on the beam centerline into the solid finite element parameter space.
For element-based integration, the evaluation time for the coupling terms is more or less proportional to the number of Gauss-points used.
Since the integrand contains kinks, a relatively high number of Gauss-points is necessary to obtain a sufficiently accurate numerical integration.
On the other hand, segment-based integration requires calculation of the intersections of the beam elements with the solid surfaces.
This intersection operation also requires the solution of local nonlinear systems.
The total number of intersections, which have to be calculated, depends on the mesh configuration and cannot be quantified in a general manner.
The advantage of the segment-based integration is that the integrands over a segment are smooth, see the left part of Figure~\ref{fig:discret_discontinuities}, and an acceptable integration error can be obtained with a reasonable number of Gauss-points.
Unless stated otherwise, all the examples in this work use 6 Gauss-points per integration segment.
A direct comparison of the two integration schemes regarding evaluation time is difficult, as the times depend on the mesh configuration of the individual problem.
In \cite{Farah2015}, an elaborate comparison of different numerical integration algorithms for mortar methods is given.
It should be stated that, in general, due to the non-polynomial integrand in $\M$ both integration schemes cannot integrate $\M$ exactly.
Nevertheless, the segment-based integration has clear advantages: the accuracy of its numerical integration is independent of the beam-to-solid element length ratio and a higher accuracy can be achieved with the same global number of Gauss points.

%% file: tex/examples.tex
\section{Examples}
\label{sec:ex}

The following examples are chosen to evaluate the different \btsvc methods proposed in this work, \cf Table~\ref{tbl:ex_coupling_mehods}, and to demonstrate their accuracy and robustness for the simulation of challenging engineering applications.
All simulations are performed with our in-house parallel multi-physics research code BACI.
\begin{table*}
	\centering
	\caption{Listing of the different coupling methods investigated in this section.}
	\label{tbl:ex_coupling_mehods}
	\begin{tabular}{llll}
	\toprule
		coupling discretization & coupling type & Lagrange multiplier shape function & numerical integration   \\
	\midrule
		\gpts                   & 1D-3D         & --                                 & element + segment based \\
		                        & 2D-3D         & --                                 & element based           \\
	\midrule
		mortar                  & 1D-3D         & linear                             & element + segment based \\
		                        &               & quadratic                          & element + segment based \\
		                        &               & cubic                              & element + segment based \\
	\bottomrule
	\end{tabular}
\end{table*}

\subsection{Patch Tests}
\label{sec:ex_patch}
Patch tests are a well-established tool to investigate the consistency of finite element formulations.
In the realm of solid mechanics, they are typically used to check that the finite element method is able to exactly represent a constant stress state in the patch.
For \sts mesh tying this is for example shown in \cite{Puso2004}, where a constant traction across non-conforming interfaces can exactly be represented via a mortar mesh finite element approach.
However, the choice of a suitable patch test for \btsvc is not straightforward, as this is technically not a discretization of two domains with non-matching meshes at the interfaces, but rather an embedded problem of a 1D curve (the beam) lying inside a 3D volume (the solid).

\subsubsection{Beams inside a solid volume}
\begin{figure}
\centering
\includegraphics[scale=1]{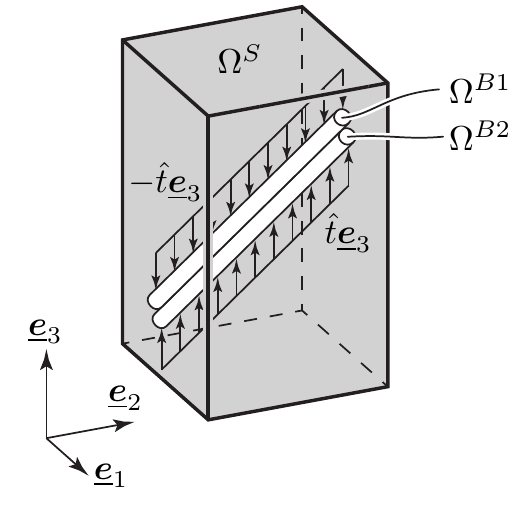}
\caption{Problem setup for the first patch test.}
\label{fig:ex_patch_problem}
\end{figure}

Figure~\ref{fig:ex_patch_problem} shows the first patch test presented here.
It consists of a solid cuboid $\OmegaSolid$ with two embedded straight beams B1 and B2, where $\OmegaBeam1$ and $\OmegaBeam2$, the domains of both beams, occupy the same spatial domain.
No surface loads or body forces are applied on the solid, while constant line loads with a magnitude $\load$ act in opposite directions $\pm \ez$ on the beams.
Therefore, the opposing loads on the two beams cancel each other out and in sum the two beams transfer no loads to the solid.
This gives the trivial solution for the solid displacement field $\uSolid = \tns{0}$ and the constant solution $-\uBeam1 = \uBeam2 = \ez \load  / \varepsilon$ for the beam displacements, where $\varepsilon$ is the contact stiffness between the beams.
This patch test uses the proposed \btsvc method to couple both beams to the solid.
By doing so, all interactions between the beams are transfered via the solid domain, resulting in a patch test like problem for \btsvc.
This test case will be used to assess the influence of discretization and integration error on the performance of the proposed \btsvc method.

The dimensions of the cube are $\unit[1]{m} \times \unit[1]{m} \times \unit[2]{m}$ and a compressible Neo-Hookean material law with Young's modulus $E = \unit[10]{N/m^2}$ and Possion's ratio $\nu = 0.3$ is employed as constitutive model.
The penalty stiffness of the \btsvc is $\pen = \unit[10^4]{N/m^2}$.
Both beams align along the space diagonal of the cuboid and have a length of $\unit[0.7 \sqrt{5}]{m}$.
Their \cs{s} are circular with a radius of $\unit[0.05]{m}$ and the constitutive parameters are $E=\unit[100]{N/m^2}, \nu=0$.
The solid is discretized with $4\times 4\times 7$ eight-noded, first order hexahedral elements (\hex8).
\sr beam elements are used to represent both beams $B1$ and $B2$, which are discretized with 5 and 7 equidistant elements, respectively.
Mortar coupling is applied between the beam centerline and the solid, with a linear interpolation of the Lagrange multiplier field $\lagrange$ along the beam elements.
To circumvent numerical problems in the solution of the resulting linear system of equations, the solid is constrained such that all six rigid body modes of the system are eliminated.
Additionally, any rotation of the first nodes of the two beams is constrained to prevent a rigid body rotation of the beams around their axes.
The magnitude of the line loads on the beams is $\load = \unit[5]{N/m}$.

\begin{figure*}
\centering
\subfigure[]{\includegraphics[resolution=300]{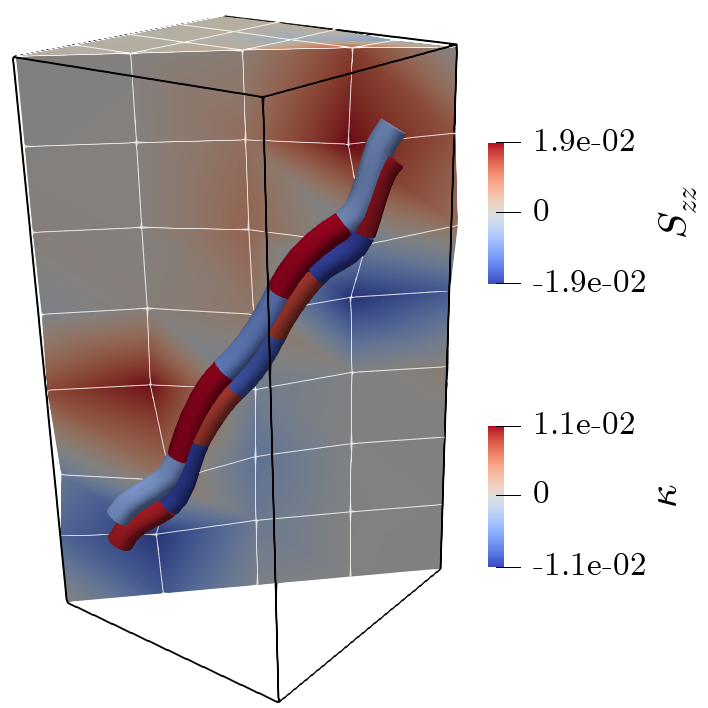}\label{fig:ex_patch_result_straight_element_based}}
\hfil
\subfigure[]{\includegraphics[resolution=300]{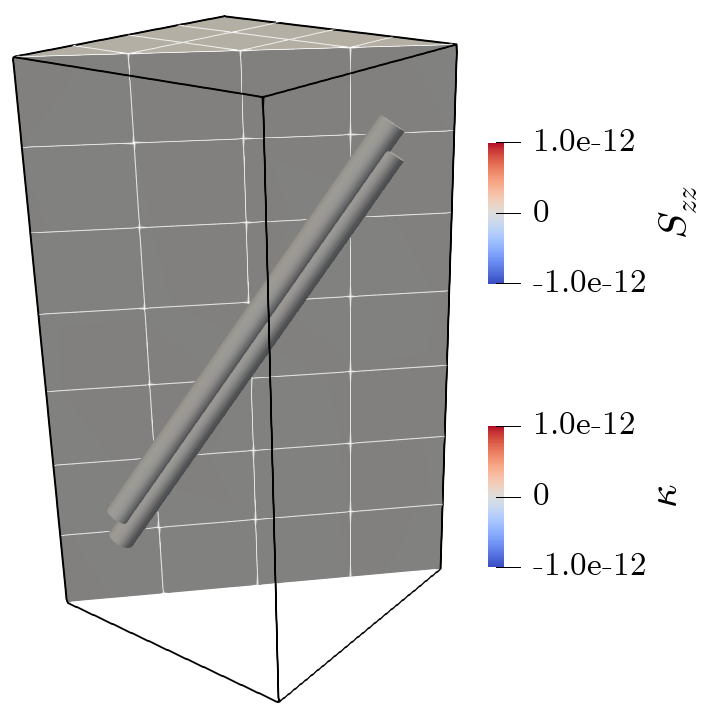}\label{fig:ex_patch_result_straight_segment_based}}
\hfil
\caption{
	Patch test with overlapping straight beams -- element-based integration \subref{fig:ex_patch_result_straight_element_based} and segment-based integration \subref{fig:ex_patch_result_straight_segment_based}.
    The second Piola-Kirchhoff stress $S_{zz}$ is shown in the solid and the curvature $\kappa$ at the middle of each beam element.
    Displacements of beams and solid are scaled with a factor of $100$.
	}
\end{figure*}

For the given geometry, there is no discretization error, since the chosen shape functions for the beams and the solid are able to exactly represent both geometry and numerical solution.
To assess the numerical integration error, the problem is solved once with element-based integration of the mortar coupling terms and once with segment-based integration.
Figure~\ref{fig:ex_patch_result_straight_element_based} shows the result obtained with element-based integration and 6 Gauss points per beam element.
Clearly, the solution is not exact, as the solid is not stress-free and the deformation of the beams is not constant, thus resulting in non-vanishing curvatures along the beams.
Figure~\ref{fig:ex_patch_result_straight_segment_based} shows the results obtained with segment-based integration of the mortar coupling terms, where each segment is integrated with 6 Gauss points.
In this case, the numerical results exactly match the analytical solution up to machine precision, which confirms the vanishing integration error for segment-based integration.

\begin{figure*}
	\centering
	\subfigure[]{\includegraphics[resolution=300]{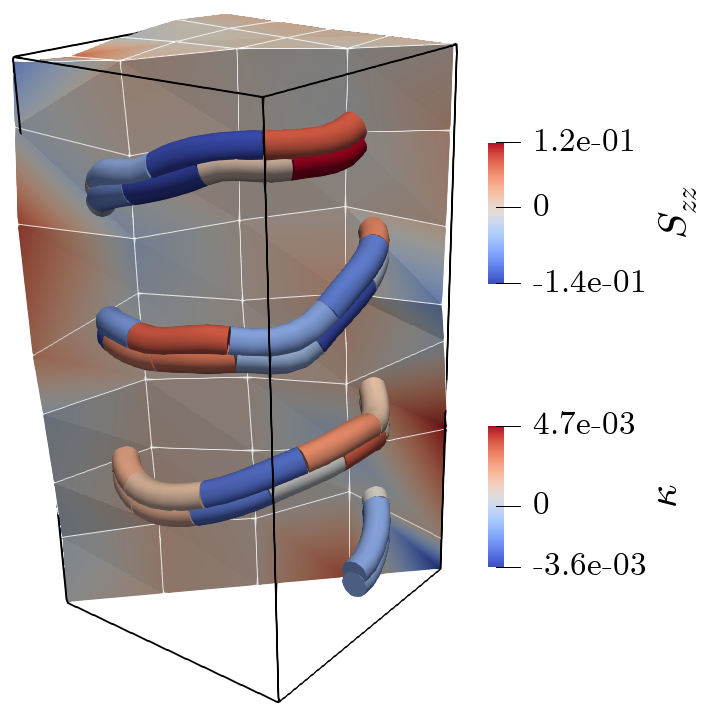}\label{fig:ex_patch_result_helix_element_based}}
	\hfil
	\subfigure[]{\includegraphics[resolution=300]{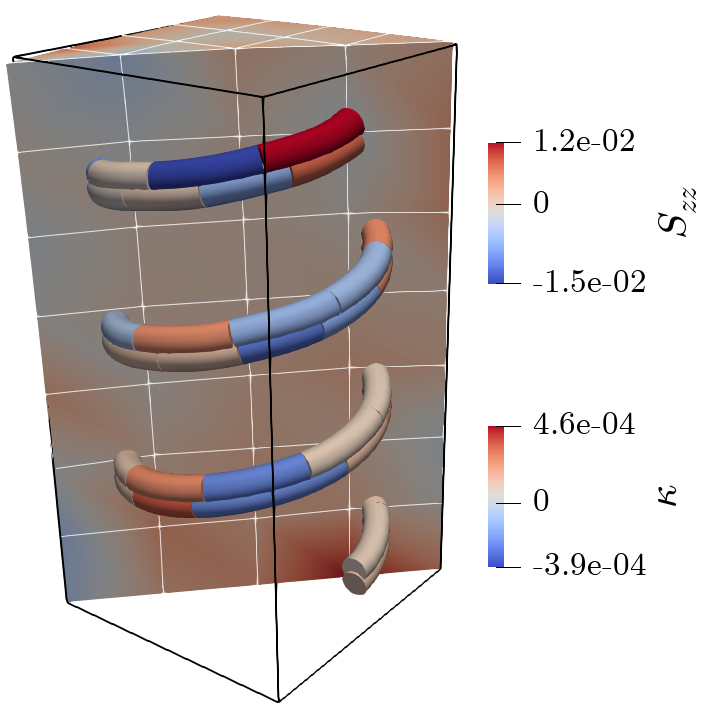}\label{fig:ex_patch_result_helix_segment_based}}
	\hfil
	\caption{
		Patch test with overlapping helix-shaped beams -- element-based integration \subref{fig:ex_patch_result_helix_element_based} and segment-based integration \subref{fig:ex_patch_result_helix_segment_based}.
		The second Piola-Kirchhoff stress $S_{zz}$ is shown in the solid and the curvature $\kappa$ at the middle of each beam element.
		Displacements of beams and solid are scaled with a factor of $50$.
	}
\end{figure*}

A second patch test is set up similar to the first one, with the straight beams being replaced by two helix-shaped beams.
The helix has the following geometrical parameters: A radius of $\unit[0.45]{m}$, three turns with a pitch of $\unit[9/5]{m}$ and a right handed screw type.
In this case, the beams B1 and B2 are discretized with 23 and 31 elements, respectively.
The employed $C^1$-continuous Hermite polynomials used for the beam centerline interpolation can not represent the helix geometry exactly, which results in two slightly different geometries of the beams and different arc lengths of the two helices, thus introducing a discretization error.
In order for the two beams to be in equilibrium, the load $\load$ on beam B2 is scaled with a factor of $0.999318$, to correct for the different beam lengths.
In this case, the beams can not perform a rigid body motion when coupled to the solid.
Therefore, only the six rigid body modes of the solid are constrained.
All other parameters are equal to the previously described patch test.
Figure~\ref{fig:ex_patch_result_helix_element_based} shows the results with element-based integration of the mortar coupling terms.
Similar to the previous scenario, one can see non-vanishing stresses in the solid and curvature oscillations in the beams.
In this case, also the result with segment-based integration, shown in Figure~\ref{fig:ex_patch_result_helix_segment_based}, does not match the analytical results up to machine precision, because of the previously described discretization error.
However, when comparing the quantitative results, one can see that the influence of the numerical integration error for element-based integration is about one order of magnitude larger than the discretization error, which confirms that element-based integration introduces a significant additional integration error.

The presented results were all calculated with our mortar \btsvc approach and first-order interpolation of the Lagrange multipliers.
Quantitatively, the results change only slightly if a \gpts (1D-3D) approach is used or if a different interpolation scheme for the Lagrange multipliers is applied.
Therefore, the conclusions obtained from the shown examples, \ie the importance of an accurate numerical integration of the \btsvc terms and the superiority of segment- over element-based integration, can be applied to all aforementioned cases.

\subsubsection{Strong discontinuities in \btsvc}

\begin{figure*}
	\centering
	\subfigure[]{\includegraphics[resolution=300]{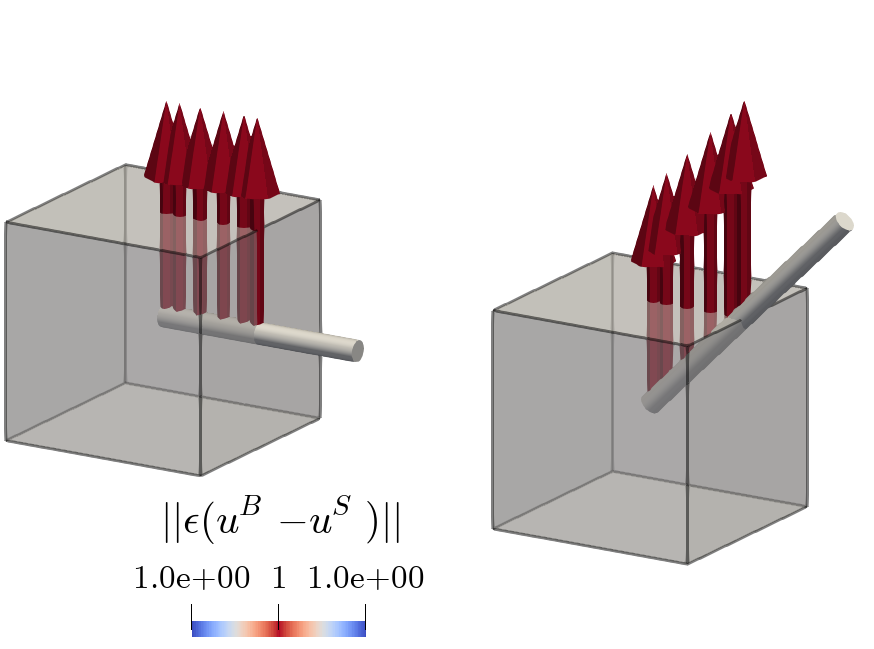}\label{fig:ex_patch_stick_out_gp}}
	\hfill
	\subfigure[]{\includegraphics[resolution=300]{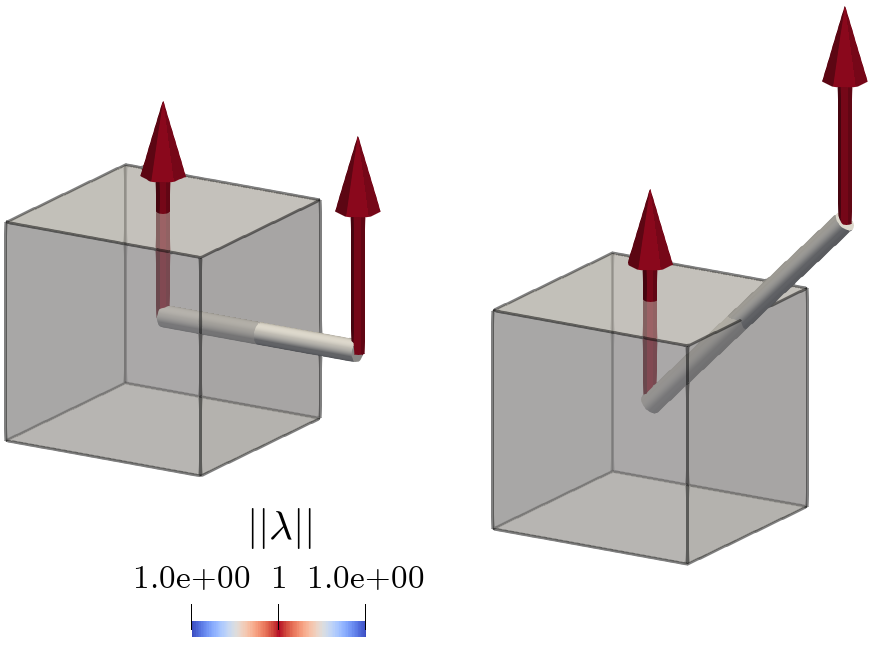}\label{fig:ex_patch_stick_out_mortar}}
	\caption{
		Strong discontinuities in \btsvc{} -- Gauss-point-to-segment approach with the negative coupling forces at the integration points \subref{fig:ex_patch_stick_out_gp} and mortar coupling with the negative discrete Lagrange multiplier traction vectors \subref{fig:ex_patch_stick_out_mortar}.
	}
	\label{fig:ex_patch_stick_out}
\end{figure*}

To check the ability of the proposed methods to handle strong discontinuities, \ie a beam sticking out of a solid domain, two more patch tests are introduced.
Both problems consist of a solid cube and a straight beam which starts inside of the cube and ends outside of it.
In the first case, the beam intersects a face of the solid, in the second one it intersects an edge.
All solid degrees of freedom are constrained and a constant line load $-\load \ez = -\unit[1]{N/m} \, \ez$ is applied only to the part of the beam inside of the cube.
Similar to the previous patch test the analytical solution for the beam displacement is $\uBeam{} = - \load  / \varepsilon \, \ez$.
Furthermore, the beam can only be in equilibrium if the coupling interface traction is $\lagrange = \load \ez = \unit[1]{N/m} \, \ez$.
For reasons of simplicity, the solid cube is discretized with a single \hex8 element, the beam with a single \sr element.
In this case, segment- and element-based integration are identical to each other, as both schemes will result in the same integration points and weights.
Segmentation has to be performed at the point where the beam exits the solid volume.
The patch tests are analyzed once with a Gauss-point-to-segment approach and once with mortar coupling using a linear interpolation of the Lagrange multipliers.
Figure~\ref{fig:ex_patch_stick_out} shows the results.
For the Gauss-point-to-segment method, the coupling forces at the integration points are illustrated and it can be observed that they are exact up to machine precision.
The same holds true for the Lagrange multiplier interface tractions in the mortar case.
The discrete Lagrange multipliers should not be confused with discrete nodal loads on the beam element, as the Lagrange multiplier field is only integrated on the beam segment that resides inside the solid.
This underlines the importance of segmentation at solid surfaces, \ie proper treatment of strong discontinuities.

\subsection{Spatial Convergence}
\label{sec:ex_spatial_convergence}

\begin{figure}
	\centering
	\includegraphics[scale=1]{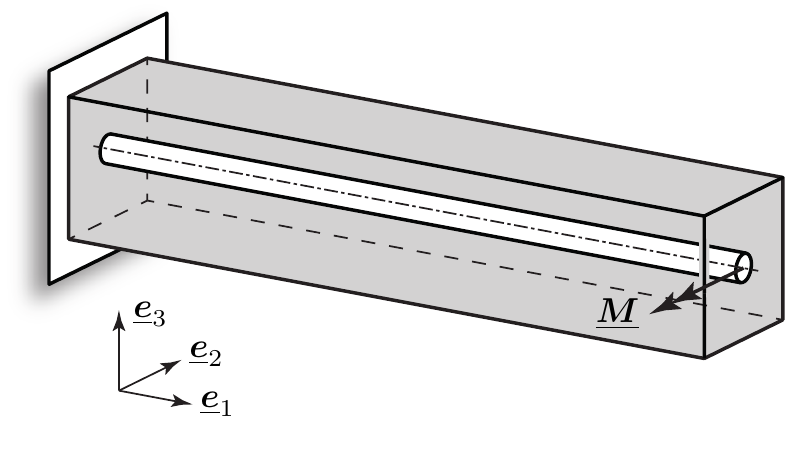}
	\caption{
		Convergence test case for \btsvc -- Problem setup of a coupled beam and solid structure.
	}
	\label{fig:ex_convergence}
\end{figure}

The following numerical example investigates the spatial convergence properties of the \btsvc method as well the validity of the evaluation of the coupling terms along the beam centerline instead of the beam surface, \ie our fundamental 1D-3D modeling assumption.
The considered problem is shown in Figure~\ref{fig:ex_convergence}.
It consists of a solid block with the dimensions $\unit[5]{m} \times \unit[1]{m} \times \unit[1]{m}$ and a hyperelastic Saint Venant–Kirchhoff material model ($E=\unit[10]{N/m^2}$, $\nu = 0.0$).
Embedded inside the solid block is a rod with the length $\unit[5]{m}$.
The beam is modeled as a \tf beam ($E=\unit[4346]{N/m^2}$, $\nu = 0$) with circular cross-section (radius $r=\unit[0.125]{m}$).
The parameters are chosen such that the rod and solid have the same bending stiffness around the $\ey$ and $\ez$ axes.
At the left end surface of the solid block, displacements are fixed as are the rod displacements and rotations.
At the right end, the rod is loaded with a moment $\tns{M} = - \unit[0.025]{Nm}\, \ey$.
No external loads are applied to the solid block.

\begin{figure}
	\centering
	\includegraphics[scale=1]{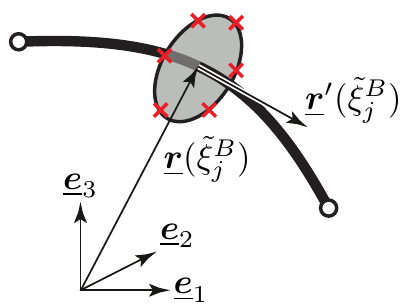}
	\caption{
		Illustration of the discrete coupling points for 2D-3D coupling along a single \cs.
	}
	\label{fig:ex_cylinder_integration}
\end{figure}

The spatial convergence behavior of the different coupling methods will be analyzed with respect to the $\L2$ displacement error
\begin{multline}
\error =
\frac{1}{V_0} \sqrt{\intsolid{ \norm{ \usolidh - \usolidhref }^2 }}
\\
+ \frac{1}{\beamlength} \sqrt{ \intbeamcenterlineDomain{ \norm{ \ubeamrh - \ubeamrhref }^2}}.
\end{multline}
Here, $V_0 = \unit[5]{m^3}$ is the solid volume in the reference configuration and $\beamlength = \unit[5]{m}$ is the reference length of the beam.
The error is computed relative to a reference finite element solution obtained with 2D-3D (surface-to-volume) coupling.
The 2D-3D coupling is realized with a \gpts approach as illustrated in Figure~\ref{fig:ex_cylinder_integration}.
At each Gauss-Legendre point $\tilde{\xi}_j^B$ along the beam centerline, multiple equally spaced coupling points (illustrated with the symbol '$\times$' in Figure~\ref{fig:ex_cylinder_integration}) are inserted along the circumference of the corresponding \cs.
The coupling points are constrained to the circumference of the \cs.
All coupling points along a single \cs ran rotate around $\rbeam'(\tilde{\xi}_j^B)$, \ie the resulting moment around $\rbeam'(\tilde{\xi}_j^B)$ vanishes, which corresponds to the kinematic assumptions of the \tf beam theory.
The coupling points are tied to the underlying solid mesh via a linear penalty constraint.
In all of the following results obtained with 2D-3D coupling, 6 Gauss-Legendre point in axial direction and 128 integration points in circumferential direction are used.
This ensures a sufficiently accurate numerical evaluation of the 2D-3D coupling terms in order to hold as reference solution, and the chosen penalty parameter does not lead to unwanted stiffening effects.

\begin{figure}
	\centering
		\includegraphics[scale=1]{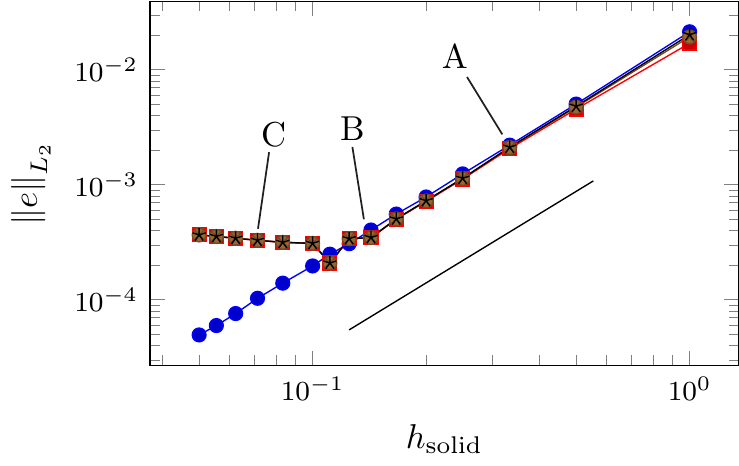}
		\includegraphics[scale=1]{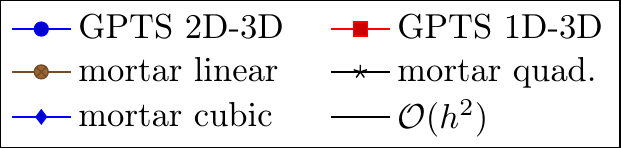}
		\\
		\vspace{0.5cm}
		\begin{tabular}{ccc}
		A & B & C \\
		\includegraphics[scale=1]{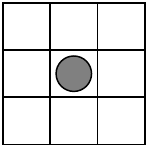}
		&
		\includegraphics[scale=1]{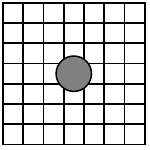}
		&
		\includegraphics[scale=1]{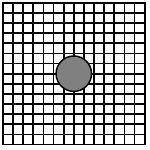}
		\end{tabular}
	\caption{
		Spatial convergence plot for different coupling methods, with solid element size compared to beam \cs diameter at certain data points.
		}
	\label{fig:ex_convergence_error}
\end{figure}

\begin{figure}
	\centering
	\subfigure[]{\includegraphics[resolution=300]{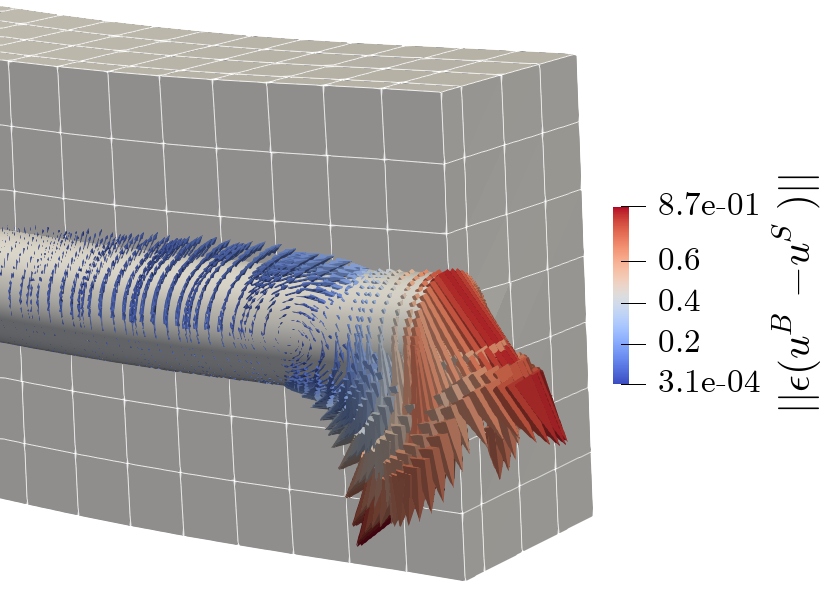}\label{fig:ex_convergence_forces_full}}
	\hfil
	\subfigure[]{\includegraphics[resolution=300]{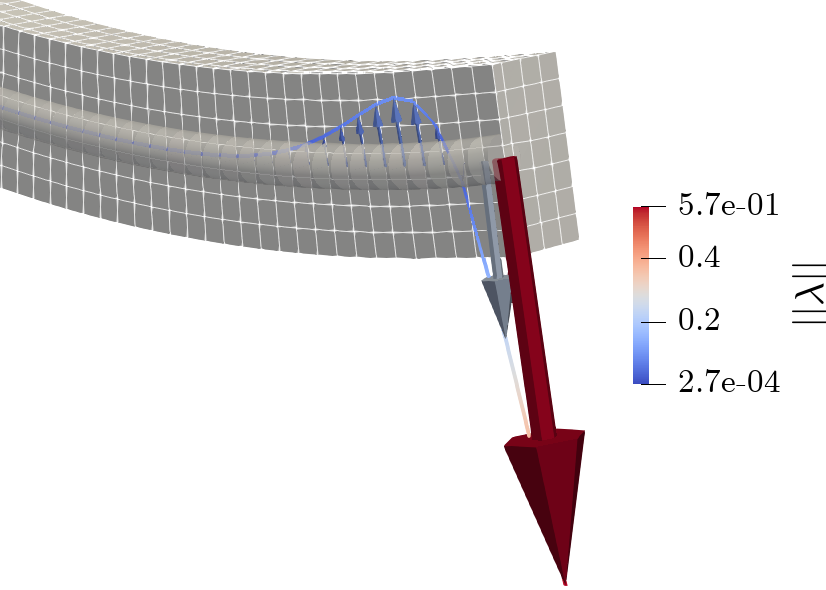}\label{fig:ex_convergence_forces_centerline}}
	\caption{Negative coupling forces for 2D-3D coupling \subref{fig:ex_convergence_forces_full} and the negative Lagrange multiplier field for 1D-3D mortar (linear interpolation) coupling \subref{fig:ex_convergence_forces_centerline}. The shown plots are for $h_{\text{solid}} = \unit[0.14]{m}$.}
	\label{fig:ex_convergence_forces}
\end{figure}

The solid block is meshed with first-order \hex8 solid elements, with an element size $h_{\text{solid}}$.
The rod is discretized with \tf beam finite elements with a length of $h_{\text{beam}} = 2.5 h_{\text{solid}}$.
The penalty parameter for all 1D-3D coupling methods is $\unit[100]{N/m^2}$, for 2D-3D \gpts coupling it is $\unit[100]{N/m^3}$.
Additionally to the previously described integration rule for 2D-3D coupling, all 1D-3D coupling schemes in this example are evaluated with segment-based integration and 6 Gauss points per segment.
These values are chosen according to Section~\ref{sec:ex_penalty_convergence} in order to avoid unwanted contact locking effects.
For models purely consisting of either first-order solid or third-order beam elements, the expected convergence rate of the $\L2$-error is $\order{h^2}$ and $\order{h^4}$, respectively.
The expected convergence rate for the coupled problem is thus the lower of the two, \ie $\order{h^2}$.
Figure~\ref{fig:ex_convergence_error} shows the convergence plot of the coupled structure with different coupling methods.
The 2D-3D \gpts coupling scheme exhibits the expected convergence rate of $\order{h^2}$ for the entire dataset.
All 1D-3D coupling schemes behave very similar to each other.
For coarse meshes, the expected optimal convergence order $\order{h^2}$ can be observed.
At around $h_{\text{solid}} = \unit[0.12]{m}$ the convergence behavior of all 1D-3D coupling methods has a kink, and for smaller element sizes the error does not decrease any further, it even slightly increases.
The bottom part of Figure~\ref{fig:ex_convergence_error} illustrates the solid mesh size compared to the beam \cs at different points in the convergence plot.
In the case of a coupling along the beam surface (2D-3D), the beam interacts with all solid elements along its surface.
For beam-to-centerline coupling (1D-3D), the beam only interacts with the solid elements along its centerline.
For finer discretizations, the influence of the different interactions becomes more evident, which materializes in the kink in the convergence plot.
This result gives rise to a very important finding, namely that our 1D-3D coupling is valid down to a certain element size, \ie up to the kink in the convergence plot.
Exemplarily, the beam tip displacement of the 2D-3D reference solution is $\unit[0.19009]{m}$. With our 1D-3D \btsvc method the tip displacement at the kink in the convergence plot (mesh B from Figure~\ref{fig:ex_convergence_error}) is $\unit[0.18895]{m}$, which amounts to a relative error of approximately $0.5\%$.
The coupling interactions of the 2D-3D and 1D-3D schemes are shown in Figure~\ref{fig:ex_convergence_forces}.
The critical solid element size, \ie up to which the 1D-3D coupling is accurate, depends on a number of different parameters and can not be given in closed form.
However, for the problems considered in this work, \ie rather stiff beams and soft solids, a rule of thumb can be given: the solid element size should not be smaller than the beam \cs diameter.
Keeping in mind the envisaged applications, one can conclude that this does not pose any restrictions on our \btsvc methods, but is perfectly in line with their modeling goal.

\begin{figure}
	\centering
	\includegraphics[scale=1]{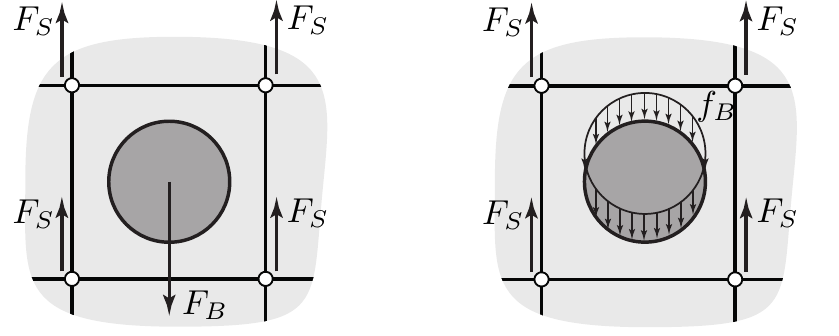}
	\caption{Plane coupling problem of a single fiber with a solid finite element mesh -- 1D-3D coupling (left), 2D-3D coupling (right).} 
	\label{fig:ex_convergence_simple}
\end{figure}

\begin{remark}
Consider a plane problem of a beam \cs coupled with a solid finite element mesh as depicted in Figure~\ref{fig:ex_convergence_simple}.
As long as the \cs fully lies within a single solid element, \ie the \cs diameter is smaller than the solid element size, the resulting nodal forces on the solid nodes should be independent of the used coupling scheme -- as long as the resultants of the 1D-3D and 2D-3D coupling are equivalent.
Obviously this is an idealized setting, but this still underlines and nicely illustrates the validity of our 1D-3D coupling approach down to a solid element size of about the \cs diameter.
\end{remark}


\subsection{Influence of the penalty parameter}
\label{sec:ex_penalty_convergence}

\begin{figure*}
	\centering
	\newcommand{\heightfig}{6cm}
	\includegraphics[scale=1]{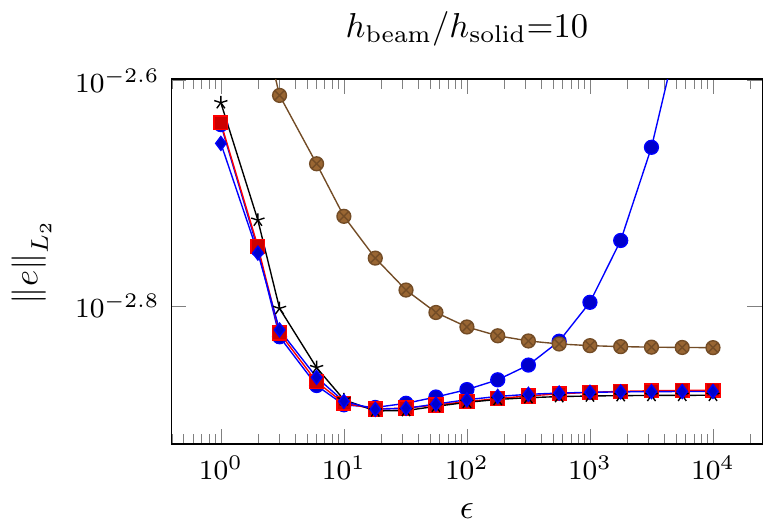}
	\hspace{7mm}
	\includegraphics[scale=1]{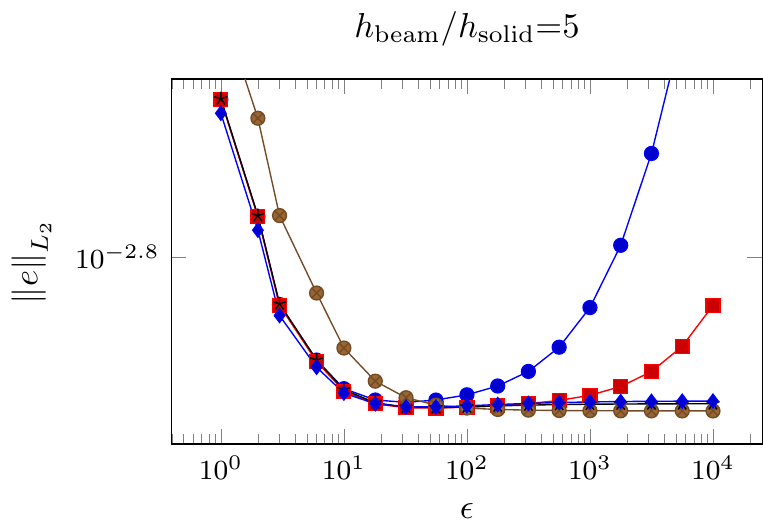}
	\\
	\includegraphics[scale=1]{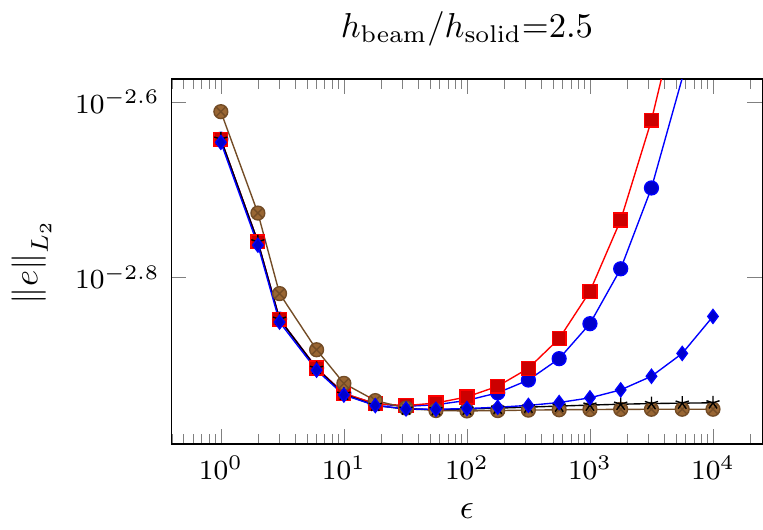}
	\hspace{7mm}
	\includegraphics[scale=1]{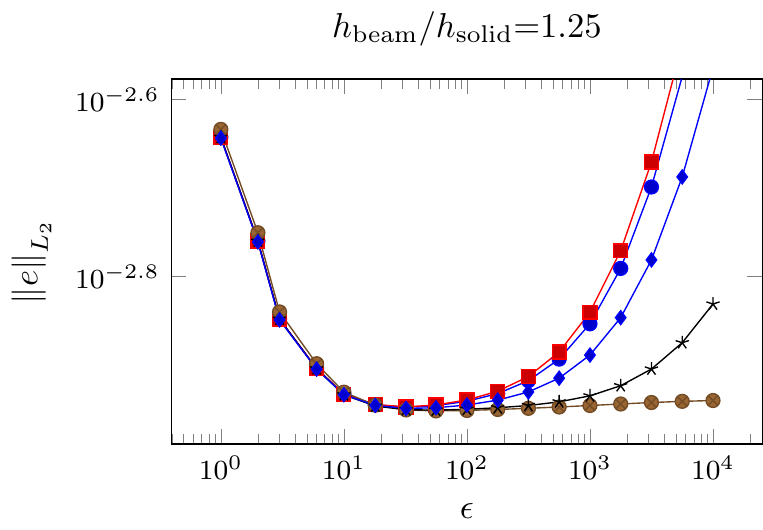}
	\\
	\includegraphics[scale=1]{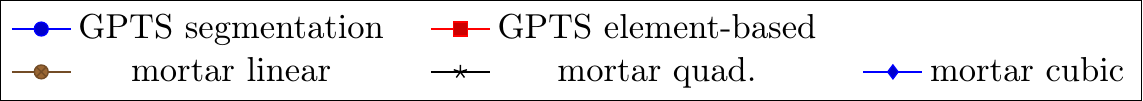}
	\caption{$\L2$-error for different parameter combinations and \btsvc schemes.}
	\label{fig:ex_convergence_penalty}
\end{figure*}

In this example, the influence of the penalty parameter as well as the beam-to-solid element length ratio is investigated.
The analyzed problem is the same as in Section~\ref{sec:ex_spatial_convergence}, now with a fixed solid element length $h_{\text{solid}}=\unit[0.25]{m}$.
The model is simulated with different penalty parameters and beam-to-solid element length ratios.
To quantify the differences between results obtained with different parameters, the $\L2$-errors relative to the same reference solution as used in Section~\ref{sec:ex_spatial_convergence} are compared.
The results are shown in Figure~\ref{fig:ex_convergence_penalty}.
Each of the four plots represents a fixed beam-to-solid element length ratio.
The penalty parameter is plotted on the abscissa.
The line style identifies the employed coupling scheme.
For both element and segment-based integration, 6 integration points are used per element and segment, respectively.
The desired behavior for an increasing penalty parameter is a convergence towards the exact fulfillment of the constraint equations, \ie the solution of \eqref{eq:discret_global_system}.
In the presented plots, this corresponds to a horizontal line for high penalty parameters.
For all element length ratios, the \gpts scheme with segment-based integration exhibits an increasing error for increasing penalty parameters.
The \gpts scheme with element-based integration behaves better for high element length ratios, but as the beam length gets closer to the solid element size, the same behavior can be observed.
This effect is sometimes referred to as contact locking and occurs due to an over-constraining of the system, \ie too many discrete coupling constraints are enforced, and as a result, the coupling discretization becomes too stiff.
Mathematically, this is related to a violation of the discrete inf-sup condition \cite{Boffi2013}.
This effect is especially distinct for \gpts schemes, where each Gauss point represents three coupling constraints, \ie the number of discrete coupling constraints depends on the integration scheme used.
A smaller number of Gauss points can usually improve the contact locking properties for \gpts schemes, but this in turn can lead to the non-fulfillment of the patch tests given in Section~\ref{sec:ex_patch}.
The \btsvc mortar schemes behave better: for element length ratios of 10 and 5 no locking can be observed at all.
For smaller element length ratios the schemes with quadratic and cubic interpolation also show signs of contact locking.
Linear interpolations of the Lagrange multipliers do not show such behavior for the considered element length ratios.
By using a lower order interpolation of the Lagrange multipliers, the number of constraints is reduced, which explains the better behavior of the lower-order Lagrange multiplier interpolations regarding contact locking.
The employed numerical integration scheme does not affect the contact locking behavior of mortar \btsvc methods, as the number of coupling constraints is independent of the number of Gauss points used.

The results show that a \gpts-based coupling discretization tends to be prone to spurious contact locking effects.
A linear interpolation of the Lagrange multipliers within a mortar-based coupling discretization, as suggested in this contribution, is the most robust coupling scheme regarding the choice of the penalty parameter.

\subsection{Fiber-Reinforced Composite Plate}
\label{sec:ex_composite_plate}

\begin{figure}
	\centering
    \includegraphics[scale=1]{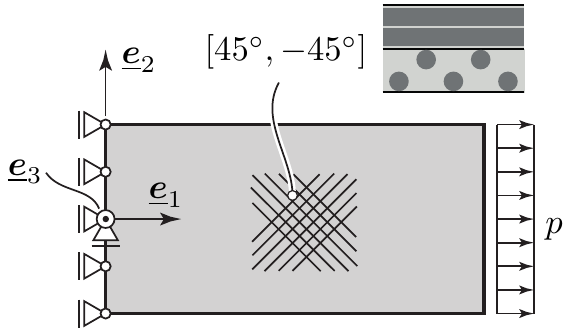}
	\caption{Problem setup of the fiber-reinforced composite plate.}
	\label{fig:ex_composite_plate_problem}
\end{figure}

In this example, a fiber-reinforced composite plate is modeled with the proposed \btsvc method and the results are compared to a homogenized approach.
Figure~\ref{fig:ex_composite_plate_problem} shows the problem setup of a two-layer composite plate.
The plate has a length and width of $\unit[2]{m}$ and $\unit[1]{m}$, respectively.
The layer buildup is asymmetric: it consists of two layers with fiber directions of $45^\circ$ and $-45^\circ$, each with a thickness of $\unit[0.02]{m}$.
A hyperelastic Saint Venant–Kirchhoff material model ($E=\unit[10]{N/m^2}$, $\nu = 0.3$) is used to model the matrix material.
The fibers are modeled as \tf beams ($E=\unit[1000]{N/m^2}$, $\nu = 0$) with circular cross-sections (radius $r=\unit[0.045]{m}$).
Figure~\ref{fig:ex_composite_plate_problem} shows the fiber placement in the layers, which results in a fiber volume ratio of $0.25$.
At one of its short ends the plate is clamped in $\ex$ and $\ez$ direction, and a surface Neumann load $p$ in $\ex$ direction of $\unit[2.5]{N/m^2}$ is applied at the other short end.
The matrix is modeled with $288$ eight-noded solid-shell elements \cite{Bischoff1997, Vu-Quoc2003a} and the fibers with $1498$ \tf beam elements, respectively.
On average, the beam-to-solid element length ratio is about $2.5$.
Mortar coupling with linear interpolation of the Lagrange multiplier shape functions and a penalty parameter of $\unit[1000]{N/m}$ is used to couple the beams to the solid.
Segment-based integration is used to evaluate the coupling terms.
All boundary conditions are exclusively applied to the solid-shell elements.

\begin{figure}
	\centering
	\subfigure[]{\includegraphics[resolution=300]{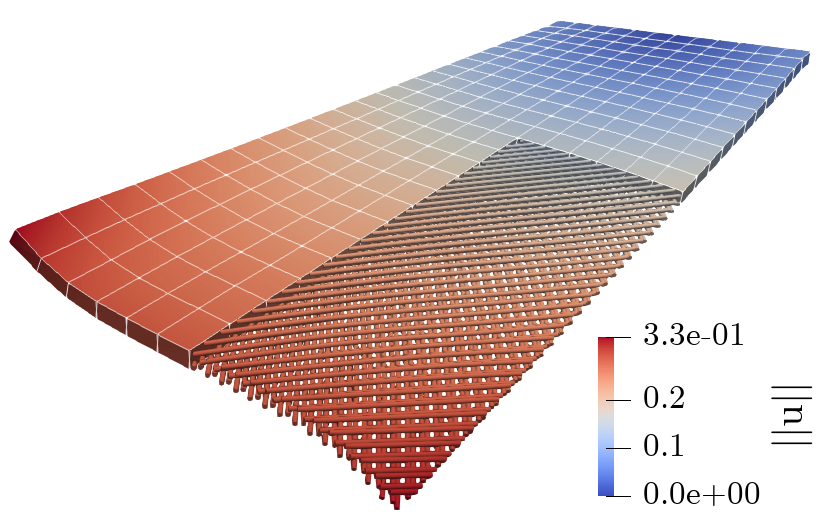}\label{fig:ex_composite_plate_deformed}}
	\subfigure[]{\includegraphics[resolution=300]{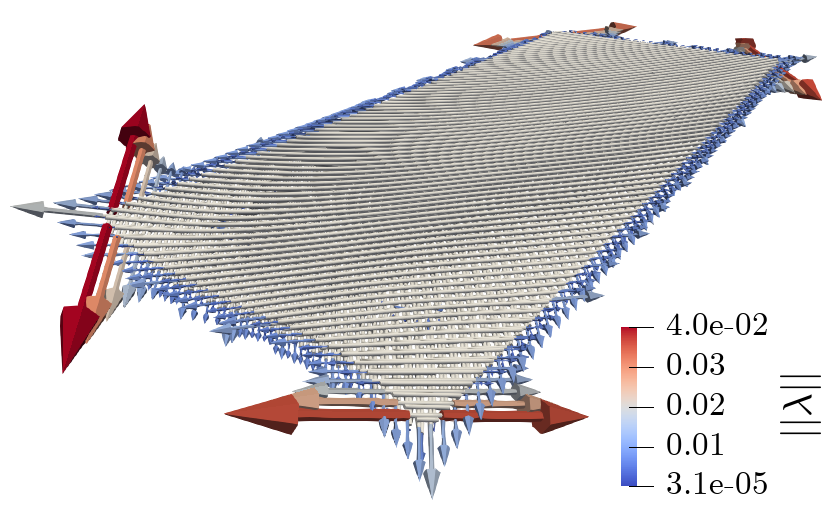}\label{fig:ex_composite_plate_force}}
	\caption{Deformed fiber-reinforced composite plate \subref{fig:ex_composite_plate_deformed} and the deformed plate with the negative discrete mortar coupling tractions \subref{fig:ex_composite_plate_force}.}
	\label{fig:ex_composite_plate}
\end{figure}

Figure~\ref{fig:ex_composite_plate_deformed} shows the deformed plate, where for illustration purposes only three quarters of the solid elements are visualized.
Due to its asymmetric layer buildup, the plate deforms out of the $\ex-\ey$ plane, even tough all applied loads and boundary conditions are exclusively in-plane.
Figure~\ref{fig:ex_composite_plate_force} shows only the beam elements and a vector plot of the negative discrete nodal values of the coupling tractions calculated with \eqref{eq:discret_penalty}.
The largest coupling tractions occur at the boundary of the plate, especially at the corners.
These coupling tractions will be used to gain insight on fiber pull-out and related composite damage phenomena in future research.
Such information cannot be obtained at all from a homogenized theory.

\begin{figure*}
	\centering
	\begin{minipage}[b]{\textwidth}
		\centering
		\raisebox{-0.5\height}{\includegraphics[scale=1]{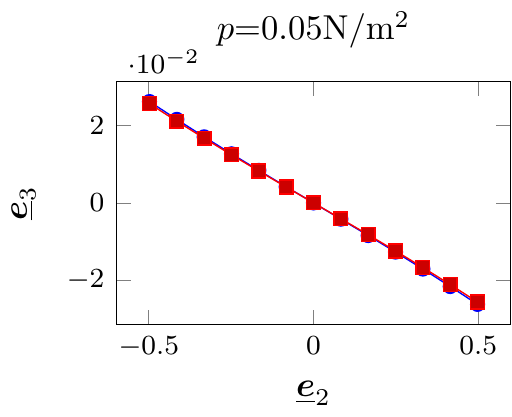}}
		\hfill
		\raisebox{-0.5\height}{\includegraphics[scale=1]{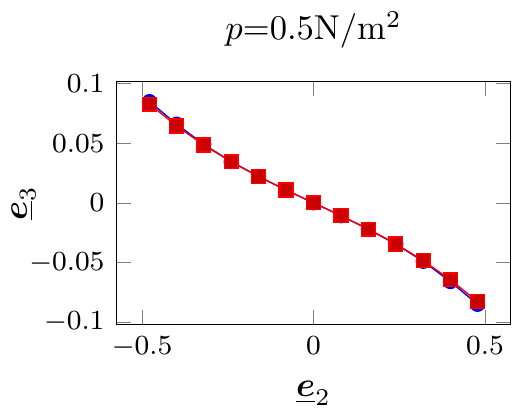}}
		\hfill
		\raisebox{-0.5\height}{\includegraphics[scale=1]{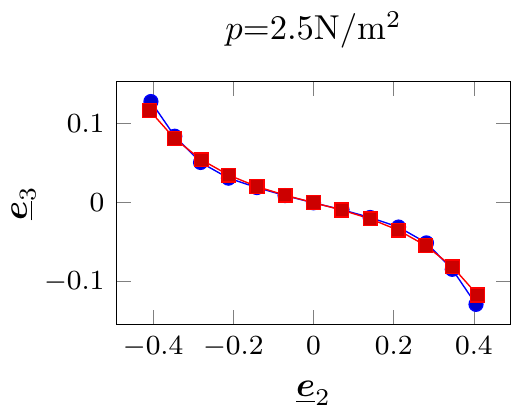}}
		\\
		\raisebox{-0.5\height}{\includegraphics[scale=1]{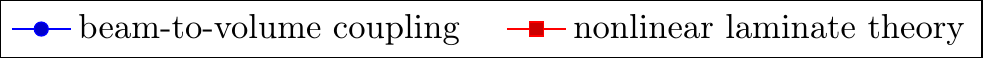}}
	\end{minipage}
	\caption{Deformed centerline of the plate at different load values and for different modeling techniques.}
	\label{fig:ex_composite_comparison}
\end{figure*}

The same plate is also modeled using a homogenized approach.
Each layer is modeled with a transversely isotropic material, this representing a homogenization of the fibers and matrix in that layer.
As is common practice, the material properties for the transversely isotropic material are calculated according to a homogenization approach for linear strains, \cf\cite{Wiedemann2007}.
For the nonlinear simulation of the plate, a combination of a purely isotropic hyperelastic material and a transversely orthotropic hyperelastic material is employed, \cf\cite{Bonet1998}.
Each layer is modeled with $288$ eight-noded solid-shell elements, thus resulting in a total of $576$ finite elements for the homogenized model.
In Figure~\ref{fig:ex_composite_comparison}, the deformations of the mid-plane at the right end ($\ex=\unit[2]{m}$) are compared to the results obtained with our new \btsvc method.
Only for larger loads, there is a tiny discrepancy between the different methods, which can be attributed to a number of factors, \eg the different strain measurements used in the beam and the homogenized solid, or small scale effects in the composite that can not be resolved by the continuum model.
Nevertheless, the results are in excellent agreement with each other, which underlines the general applicability of our \btsvc method to fiber-reinforced composites.

\begin{remark}
The presented \btsvc model of the composite plate consists of $1{,}950$ solid degrees of freedom and $10{,}992$ (\tf) beam finite element degrees of freedom.
This example can also be modeled with \kl beam elements, which yields the same numerical results up to machine precision, due to exactly vanishing torsion \cite{Meier2015}.
However, the number of beam degrees of freedom for the \kl model increases by about $30\%$ to $14{,}322$, thus justifying and encouraging the application of torsion-free beam element formulations if the underlying assumptions are met.
\end{remark}

\subsection{Fiber-Reinforced Pipe}
\begin{figure*}
	\centering
	\subfigure[]{\includegraphics[resolution=300]{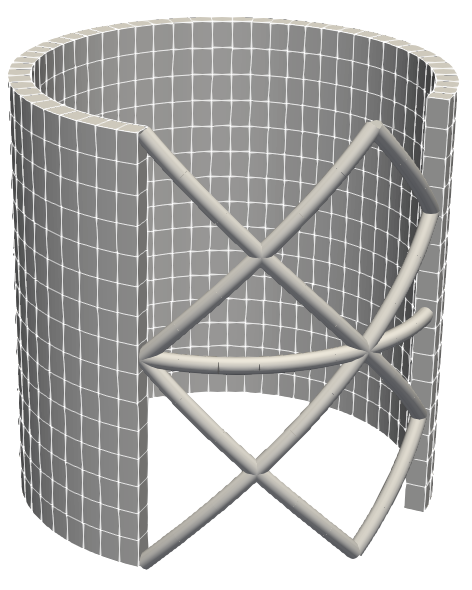}\label{fig:ex_pressure_pipe_ref}}
	\hfil
	\subfigure[]{\includegraphics[resolution=300]{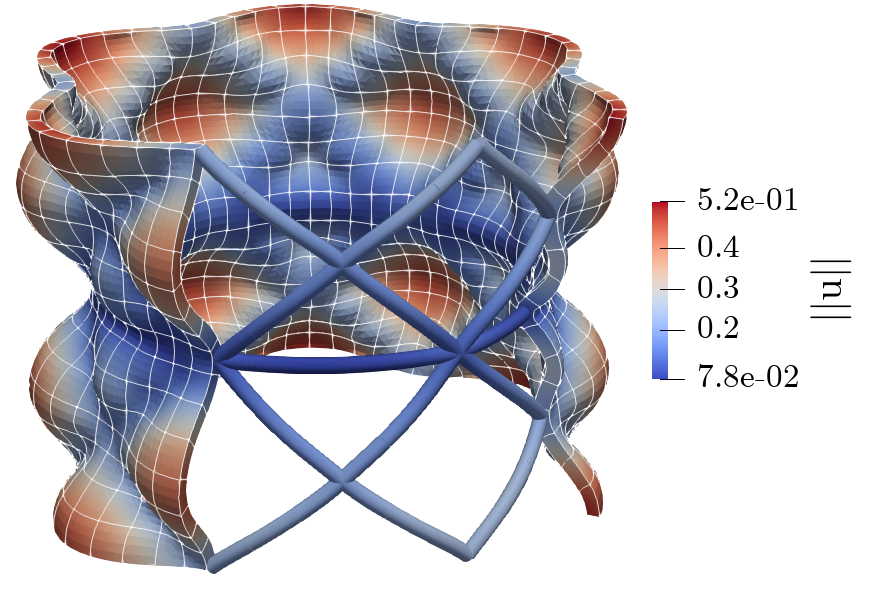}\label{fig:ex_pressure_pipe_deformed}}
	\caption{Fiber-reinforced pipe under pressure -- undeformed reference configuration \subref{fig:ex_pressure_pipe_ref} and deformed configuration \subref{fig:ex_pressure_pipe_deformed}.}
	\label{fig:ex_pressure_pipe}
\end{figure*}

The final numerical example is a fiber-reinforced pipe under pressure.
The problem setup, illustrated in Figure~\ref{fig:ex_pressure_pipe_ref}, consists of a pipe modeled  with a Neo-Hookean material law ($E=\unit[10]{N/m^2}$, $\nu = 0.3$).
The pipe is $\unit[2]{m}$ long and has an inner and outer radius of $\unit[0.9]{m}$ and $\unit[1]{m}$, respectively.
It is reinforced with \sr beams ($E=\unit[1000]{N/m^2}$, $\nu = 0$) as also shown in Figure~\ref{fig:ex_pressure_pipe_ref}.
The \cs radius of the beams is $\unit[0.04]{m}$.
The inner surface of the pipe is loaded with a Neumann surface pressure $p$ of up to $\unit[2.5]{N/m^2}$.
At the top and bottom, symmetry boundary conditions are applied to the pipe as well as to the beams.
For further symmetry reasons, only a quarter of the depicted pipe is actually simulated with the following element numbers referring to the quarter model.
Coupling between the beams and the solid is realized with our mortar approach and linear Lagrange multiplier shape functions with a penalty parameter of $\unit[1000]{N/m}$ and segment-based integration.
The pipe is discretized with $225$ \C1-continuous isogeometric solid elements (based on second-order NURBS) and $45$ \sr beam elements.

Figure~\ref{fig:ex_pressure_pipe_deformed} shows the deformed configuration of the pipe.
The expected stiffening effects of the beams onto the structure can clearly be seen.
In-between the beam reinforcements, the relatively soft pipe exhibits larger displacements.
Although only qualitative in nature, this example could obviously not be modeled with a homogenized approach and illustrates a very interesting problem class for the new \btsvc method.
Even tough all the previous derivations and examples used first-order interpolation of the solid finite elements, this example also showcases the straightforward applicability of our \btsvc method to higher-order and even \C1-continuous solid interpolations.
This allows for a coupling of beam and solid fields with the same order of interpolation continuity.

%% file: tex/conclusions.tex
\section{Conclusion}
\label{sec:conclusion}

In this work, we have proposed new modeling techniques for the coupling of 1D continua embedded inside full 3D continua.
Two different finite element-based coupling schemes have been introduced, a Gauss-point-to-segment (\gpts) and an embedded mortar-type approach.
The resulting constraint equations of both schemes are enforced via a penalty method, and in the case of the mortar-type approach, the penalty regularization was performed in a weighted node-wise manner.
For the mortar-type method, different discrete Lagrange multiplier bases were investigated.
Moreover, different numerical integration methods of the coupling terms were compared.
Several numerical experiments have been conducted to assess the behavior of the different schemes regarding the choice of the penalty parameter and the numerical integration of the coupling terms.
For relevant physical application scenarios of the \btsvc method, \ie relatively slender and stiff fibers compared to the surrounding matrix material, the validity of the fundamental modeling assumption of 1D-3D coupling has been verified, and its optimal spatial convergence behavior has been shown numerically.
Furthermore, the results underline the importance of an accurate numerical integration of the coupling terms as provided only by carefully chosen segmentation schemes.
Overall, the embedded mortar-type discretization with linear interpolation of the discrete Lagrange multiplier basis emerges as the better modeling choice due to its superior robustness regarding the choice of the penalty parameter, the beam element to solid element length ratio and its optimal spatial convergence properties.

Future work will focus on the extension of the presented \btsvc approach to \btssc as well as beam-to-solid surface contact.
Another topic of interest for further research is to make use of the improved local resolution of the numerical solution close to the fiber-matrix interface, compared to homogenized approaches, for analyzing progressive damage and failure phenomena in fiber-reinforced materials, such as fiber pull-out and the onset of delamination.